\documentclass[fleqn,usenatbib]{mnras}

\usepackage{newtxtext,newtxmath}
\usepackage[T1]{fontenc}

\newcommand{\orcit}[1]{\protect\href{https://orcid.org/#1}{\protect\includegraphics[width=8pt]{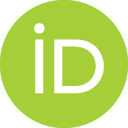}}}

\DeclareRobustCommand{\VAN}[3]{#2}
\let\VANthebibliography\thebibliography
\def\thebibliography{\DeclareRobustCommand{\VAN}[3]{##3}\VANthebibliography}

\usepackage{graphicx}	% Including figure files
\usepackage{amsmath}	% Advanced maths commands
\title[BL Lacertae at its maximum brightness levels]{The optical behaviour of BL Lacertae at its maximum brightness levels: a blend of geometry and energetics}
\author[C. M. Raiteri et al.] { 
C.~M.~Raiteri\orcit{0000-0003-1784-2784}               $^{ 1}$\thanks{E-mail:claudia.raiteri@inaf.it}, 
M.~Villata\orcit{0000-0003-1743-6946}                  $^{ 1}$,
S.~G.~Jorstad                                          $^{ 2,
 3}$,
A.~P.~Marscher                                         $^{ 2}$,
J.~A.~Acosta Pulido                                    $^{ 4,
 5}$,
 \newauthor
D.~Carosati                                            $^{ 6,
 7}$,
W.~P.~Chen                                             $^{ 8}$,
M.~D.~Joner\orcit{0000-0003-0634-8449}                 $^{ 9}$,
S.~O.~Kurtanidze                                       $^{10}$,
C.~Lorey                                               $^{11}$,
A.~Marchini                                            $^{12}$,
\newauthor
K.~Matsumoto                                           $^{13}$,
D.~O.~Mirzaqulov                                       $^{14}$,
S.~S.~Savchenko\orcit{0000-0003-4147-3851}             $^{ 3,
15,
16}$,
A.~Strigachev                                          $^{17}$,
O.~Vince                                               $^{18}$,
\newauthor
P.~Aceti                                               $^{19,
20}$,
G.~Apolonio                                            $^{ 9}$,
C.~Arena                                               $^{21}$,
A.~Arkharov                                            $^{16}$,
R.~Bachev                                              $^{17}$,
N.~Bader                                               $^{11}$,
M.~Banfi                                               $^{19}$,
\newauthor
G.~Bonnoli                                             $^{22}$,
G.~A.~Borman\orcit{0000-0002-7262-6710}                $^{23}$,
V.~Bozhilov                                            $^{24}$,
L.~F.~Brown                                            $^{25}$,
W.~Carbonell                                           $^{25}$,
M.~I.~Carnerero\orcit{0000-0001-5843-5515}             $^{ 1}$,
\newauthor
G.~Damljanovic                                         $^{18}$,
V.~Dhiman                                              $^{26,
27}$,
S.~A.~Ehgamberdiev                                     $^{14,
28}$,
D.~Elsaesser                                           $^{29,
11}$,
M.~Feige                                               $^{11}$,
\newauthor
D.~Gabellini                                           $^{30}$,
D.~Gal\'an                                             $^{ 5}$,
G.~Galli                                               $^{31}$,
H.~Gaur                                                $^{26}$,
K.~Gazeas\orcit{0000-0002-8855-3923}                   $^{32}$,
T.~S.~Grishina\orcit{0000-0002-3953-6676}              $^{ 3}$,
A.~C.~Gupta                                            $^{33,
34,
26}$,
\newauthor
V.~A.~Hagen-Thorn\orcit{0000-0002-6431-8590}           $^{ 3}$,
M.~K.~Hallum                                           $^{ 2}$,
M.~Hart                                                $^{ 2}$,
K.~Hasuda                                              $^{35}$,
K.~Heidemann                                           $^{11}$,
B.~Horst                                               $^{11}$,
\newauthor
W.-J.~Hou                                              $^{ 8}$,
S.~Ibryamov                                            $^{36}$,
R.~Z.~Ivanidze                                         $^{10}$,
M.~D.~Jovanovic                                        $^{18}$,
G.~N.~Kimeridze                                        $^{10}$,
S.~Kishore                                             $^{26,
37}$,
\newauthor
S.~Klimanov                                            $^{16}$,
E.~N.~Kopatskaya\orcit{0000-0001-9518-337X}            $^{ 3}$,
O.~M.~Kurtanidze                                       $^{10,
38,
39}$,
P.~Kushwaha                                            $^{40,
26}$,
D.~J.~Lane\orcit{0000-0002-6097-8719}                  $^{41}$,
\newauthor
E.~G.~Larionova\orcit{0000-0002-2471-6500}             $^{ 3}$,
S.~Leonini                                             $^{42}$,
H.~C.~Lin                                              $^{ 8}$,
K.~Mannheim                                            $^{43,
11}$,
G.~Marino                                              $^{21,
44}$,
M.~Minev                                               $^{24,
17}$,
\newauthor
A.~Modaressi                                           $^{25}$,
D.~A.~Morozova\orcit{0000-0002-9407-7804}              $^{ 3}$,
F.~Mortari                                             $^{30}$,
S.~V.~Nazarov                                          $^{45}$,
M.~G.~Nikolashvili                                     $^{10}$,
\newauthor
J.~Otero Santos                                        $^{ 4,
 5}$,
E.~Ovcharov                                            $^{24}$,
R.~Papini                                              $^{44}$,
V.~Pinter                                              $^{46,
47,
48}$,
C.~A.~Privitera                                        $^{12}$,
T.~Pursimo                                             $^{46,
47}$,
\newauthor
D.~Reinhart                                            $^{11}$,
J.~Roberts                                             $^{ 9}$,
F.~D.~Romanov\orcit{0000-0002-5268-7735}               $^{49,
50}$,
K.~Rosenlehner                                         $^{11}$,
T.~Sakamoto                                            $^{35}$,
F.~Salvaggio                                           $^{21,
44}$,
\newauthor
K.~Schoch                                              $^{11}$,
E.~Semkov                                              $^{17}$,
J.~Seufert                                             $^{11}$,
D.~Shakhovskoy                                         $^{45}$,
L.~A.~Sigua                                            $^{10}$,
C.~Singh                                               $^{25}$,
R.~Steineke                                            $^{11}$,
\newauthor
M.~Stojanovic                                          $^{18}$,
T.~Tripathi                                            $^{26,
37}$,
Y.~V.~Troitskaya\orcit{0000-0002-9907-9876}            $^{ 3}$,
I.~S.~Troitskiy\orcit{0000-0002-4218-0148}             $^{ 3}$,
A.~Tsai                                                $^{ 8}$,
A.~Valcheva                                            $^{24}$,
\newauthor
A.~A.~Vasilyev\orcit{0000-0002-8293-0214}              $^{ 3}$,
K.~Vrontaki                                            $^{32}$,
Z.~R.~Weaver                                           $^{ 2}$,
J.~H.~F.~Wooley                                        $^{ 9}$,
E.~Zaharieva                                           $^{24}$,
and A.~V.~Zhovtan                                       $^{45}$\\
{\it Affiliations are listed at the end of the paper} 
}

\date{Accepted XXX. Received YYY; in original form ZZZ}

\pubyear{2022}

\begin{document}
\label{firstpage}
\pagerange{\pageref{firstpage}--\pageref{lastpage}}
\maketitle

% Abstract of the paper
\begin{abstract}
In 2021 BL Lacertae underwent an extraordinary activity phase, which was intensively followed by the Whole Earth Blazar Telescope (WEBT) Collaboration. We present the WEBT optical data in the $BVRI$ bands acquired at 36 observatories around the world. In mid 2021 the source showed its historical maximum, with $R=11.14$. The light curves display many episodes of intraday variability, whose amplitude increases with source brightness, in agreement with a geometrical interpretation of the long-term flux behaviour. This is also supported by the long-term spectral variability, with an almost achromatic trend with brightness. In contrast, short-term variations are found to be strongly chromatic and are ascribed to energetic processes in the jet. 
We also analyse the optical polarimetric behaviour, finding evidence of a strong correlation between the intrinsic fast variations in flux density and those in polarisation degree, with a time delay of about 13 h. This suggests a common physical origin. The overall behaviour of the source can be interpreted as the result of two mechanisms: variability on time scales greater than several days is likely produced by orientation effects, while either shock waves propagating in the jet, or magnetic reconnection, possibly induced by kink instabilities in the jet, can explain variability on shorter time scales. The latter scenario could also account for the appearance of quasi-periodic oscillations, with periods from a few days to a few hours, during outbursts, when the jet is more closely aligned with our line of sight and the time scales are shortened by relativistic effects. 

\end{abstract}

% Select between one and six entries from the list of approved keywords.
% Don't make up new ones.

%%%%%%%%%%%%%%%%%%%%%%%%%%%%%%%%%%%%%%%%%%%%%%%%%%

%%%%%%%%%%%%%%%%% BODY OF PAPER %%%%%%%%%%%%%%%%%%
\section{Introduction}
Active galactic nuclei (AGN) are among the most powerful sources in the Universe. 
Their central engine is a supermassive black hole fed by an accretion disc.
Some AGN exhibit two plasma jets launched roughly perpendicularly to the accretion disc. In blazars, one relativistic jet is directed towards us, so that the jet emission undergoes Doppler boosting. This implies a series of effects, among which are an enhancement of the flux, a blue-shift of the emitted frequencies, and a shortening of the variability time scales \citep[e.g.][]{urry1995,blandford2019}.
The strength of these effects is described by the Doppler factor $\delta=[\Gamma \, (1-\beta \, \cos \theta)]^{-1}$, where $\Gamma=(1-\beta^2)^{-1/2}$ is the bulk Lorentz factor, $\beta$ is the plasma bulk velocity in units of the speed of light, and $\theta$ is the viewing angle. Therefore, the relativistic effects become stronger if the plasma velocity increases or the orientation of the jet becomes closer to the line of sight.

Because of the Doppler boosting, the variable jet emission of blazars usually dominates over the other contributions, coming from the AGN nucleus (accretion disc and emission line regions) or host galaxy stars, and is usually observed at all wavelengths, from the radio to the $\gamma$-ray band.
The study of blazar variability is a formidable tool to understand the structure of, and the physical mechanisms acting in, extragalactic jets. 

%{Blazar variability is best studied by monitoring the source behaviour in a multiwavelength framework. This allows us to recognise correlations between flux variations at different frequencies and their time delays, which can shed light on the processes producing the observed radiation and on the location of the emitting regions inside the jet.}

The spectral energy distribution (SED) of blazars in the $\log \nu F_\nu$ versus $\log \nu$ diagram shows two bumps, corresponding to low and high-energy emission. The frequency at which these bumps peak varies from source to source and even for the same source at different epochs.
The low-energy bump, extending from radio up to UV or even X-rays, is thought to be polarised synchrotron radiation produced by relativistic electrons in the magnetised plasma, whereas the high-energy bump, from X-rays to $\gamma$-rays, is likely predominantly due to inverse-Compton scattering of soft photons on the same relativistic electrons. However, the association between high-energy neutrinos of astrophysical origin revealed by neutrino detectors, and blazar jets as possible cosmic accelerators of these particles \citep[e.g.][]{ice2018a,ice2018b,giommi2020}, makes it probable that also hadronic processes are involved in the production of high-energy photons \citep[e.g.][]{boettcher2013}.

Detailed analysis of blazar variability requires continuous monitoring. To increase the light-curve sampling, in particular in the optical band, the effort of many observers around the world is needed. This was the founding idea of the Whole Earth Blazar Telescope\footnote{https://www.oato.inaf.it/blazars/webt/} (WEBT) Collaboration.
In 25 years of activity, the WEBT has analysed the multiwavelength behaviour of several blazars, proposing models to interpret it. 
One of the sources best studied by the WEBT is BL Lacertae \citep{villata2002,villata2004a,villata2004b,bach2006,papadakis2007,raiteri2009,villata2009,larionov2010,raiteri2010,raiteri2013,weaver2020,jorstad2022}, the prototype of the BL Lac blazar class.
In most of these previous works, we proposed that the long-term variability can be due to changes of the Doppler factor, most likely produced by variations of the orientation of the emitting regions with respect to the line of sight. In contrast, the short-term variability was ascribed to energetic processes occurring inside the jet. This scenario was found to be a viable explanation also for the variability observed in other blazars \citep[see e.g.][]{raiteri2017_nature, raiteri2021a,raiteri2021b}.
Moreover, in \citet{jorstad2022} the excellent sampling reached by the WEBT allowed us to recognise a phase of quasi-periodic oscillations (QPOs) of the optical flux, optical polarisation, and $\gamma$-ray flux.
This transient phenomenon was explained as caused by the development of a current-driven kink instability in the jet, which was caused by the passage of an off-axis disturbance past a recollimation shock. These oscillations, with period of about 13 hours, were detected during the first, most dramatic phase of the 2020 outburst.
In this new paper on BL Lacertae,  we present optical photometric and polarimetric data with extremely dense sampling acquired by the WEBT during the subsequent 2021-2022 observing season, which was characterised by the source achieving historic brightness levels and displaying rapid variability on multiple time scales.
We aim to investigate which are the persistent features in BL Lacertae optical variability behaviour, which features are instead transient, and what this can tell us about the source.
A long-term analysis of the multiwavelength behaviour of the source from the radio to the $\gamma$-ray band will follow in a subsequent paper (Raiteri et al., in preparation).

This paper is organised as follows: the optical multiband photometric observations are presented in Sect.~\ref{sec:phot}; in Sect.~\ref{sec:idv} the intraday variability is analysed, while Sect.~\ref{sec:spec} deals with colour indices and spectral variability. A wavelet analysis of the flux density behaviour is performed in Sect.~\ref{sec:wavelet}. The results of optical polarimetric monitoring are reported in Sect.~\ref{sec:pola}. In Sect.~\ref{sec:snake}, we discuss the twisting jet model that we have already successfully applied to explain blazar variability in previous papers, and in Sect.~\ref{sec:corre} we investigate the correlation between the flux and the degree of polarisation. Conclusions are drawn in Sect.~\ref{sec:fine}.

%Blazars are known to show short-term variability; in  particular intra-day variability \citep[IDV, e.g.][]{wagner1990,gupta2008}.

\section{Photometric observations}
\label{sec:phot}
As mentioned in the Introduction, the WEBT Collaboration has been intensively monitoring the behaviour of BL Lacertae over more than two decades, and in particular, closely following the most recent activity phase that started in 2020. The WEBT $R$-band light curve from 2020 March 1 to December 31 was published by \citet{jorstad2022}, while in this paper we present data from 2021 January 1 to 2022 February 28. In this period we collected 
24765 data in four optical bands: 4642 in $B$, 5370 in $V$, 12293 in $R$, and 2460 in $I$.

\begin{table*}
\caption{Details on the 43 optical datasets contributing to this paper.}
\label{tab:webt}
\begin{tabular}{llrcrcc}
\hline
Dataset                &  Country        & Diameter & Filters & $N_{\rm obs}$ & Symbol & Colour \\   % Autori inseriti
\hline
Abastumani             &     Georgia     &   70  & $R$    &  671 & {\LARGE $\diamond$} & dark green \\  %ok
Abbey Ridge            &     Canada      &   35  & $BVRI$ &  268 & {\large $\rhd$}     & orange \\ %ok
Aoyama Gakuin          &     Japan       &   35  & $BVRI$ &   18 & {$\square$}         & cyan \\ %ok 
ARIES                  &     India       &  104  & $BVRI$ &   19 & {$\square$}         & blue \\  %ok 
ARIES                  &     India       &  130  & $BVRI$ &   27 & {$\square$}         & green \\  %ok     
Athens$^a$             &     Greece      &   40  & $R$    &  141 & {\LARGE $\diamond$} & cyan \\ %ok
Beli Brezi             &     Bulgaria    &   20  & $VR$   &  152 & {\large $\ast$}     & blue \\
Belogradchik$^b$       &     Bulgaria    &   60  & $BVRI$ &  191 & {\large $+$}        & cyan \\  %ok  
Burke-Gaffney          &     Canada      &   61  & $VR$   &  660 & {\large $\rhd$}     & dark green\\ %ok
Catania (Arena)        &     Italy       &   20  & $BVR$  &   40 & {$\times$}          & cyan \\  %ok  
Catania (GAC)          &     Italy       &   25  & $VRI$  &   61 & {$\triangle$}       & cyan \\  %ok    
Connecticut            &     US          &   51  & $VR$   &  434 & {\LARGE $\ast$}     & grey \\  %ok    
Crimean (AP7p)         &     Russia      &   70  & $BVRI$ &  368 & {$\times$}          & magenta \\  %ok  
Crimean (ST-7)         &     Russia      &   70  & $BVRI$ &  432 & {\large $+$}        & magenta \\   %ok
Crimean (ST-7; pol)$^b$ &     Russia      &   70  & $R$    &  535 & {$\times$}          & dark green \\  %ok
Crimean (ZTSh)$^c$     &     Russia      &  260  & $R$    &  223 & {$\triangle$}       & red \\ %ok
Felizzano              &     Italy       &   20  & $R$    &   14 & {\LARGE $\ast$}     & magenta \\ %ok
GiaGa                  &     Italy       &   36  & $BVR$  &   62 & {\LARGE $\ast$}     & black \\ %ok
Haleakala (LCO)        &     US          &   40  & $VR$   &   50 & {\large $+$}        & blue \\     %ok 
Hans Haffner           &     Germany     &   50  & $BVR$  & 1254 & {\LARGE $\circ$}    & red \\  %ok    
Hypatia                &     Italy       &   25  & $R$    &  827 & {\LARGE $\diamond$} & red \\ %ok    
Lowell (LDT)           &     US          &  430  & $VR$   &   18 & {\LARGE $\circ$}    & magenta \\  %ok   
Lulin (SLT)            &     Taiwan      &   40  & $R$    &  592 & {$\times$}          & violet \\  %ok   
McDonald (LCO)         &     US          &   40  & $VR$   &  117 & {$\times$}          & blue \\   %ok   
Montarrenti            &     Italy       &   53  & $BVRI$ &  434 & {\LARGE $\circ$}    & dark green \\ %ok    
Monte San Lorenzo      &     Italy       &   53  & $R$    &  165 & {\LARGE $\circ$}    & green \\ %ok
Mt. Maidanak           &     Uzbekistan  &   60  & $BVRI$ & 2961 & {\LARGE $\diamond$} & green \\  %ok   
Osaka Kyoiku           &     Japan       &   51  & $BR$   &  569 & {$\square$}         & orange \\  %ok   
Perkins$^b$            &     US          &  180  & $BVRI$ &  824 & {\LARGE $\circ$}    & blue \\   %ok  
Roque (NOT; e2v)$^b$   &     Spain       &  256  & $BVRI$ &  124 & {\large $+$}        & green \\ %ok     
Rozhen                 &     Bulgaria    &  200  & $BVRI$ &   51 & {$\square$}         & red \\  %ok
Rozhen                 &     Bulgaria    & 50/70 & $BVRI$ &  181 & {$\times$}          & orange \\  %ok
%San Pedro Martir       &     Mexico      &   84  & $R$    &    2 & {$\times$}          & red \\ %ok
Seveso                 &     Italy       &   30  & $VR$   &   69 & {\large $+$}        & violet \\  %ok
Siena                  &     Italy       &   30  & $VRI$  & 2367 & {\LARGE $\diamond$} & blue \\    %ok 
Skinakas               &     Greece      &  130  & $BVRI$ & 1976 & {$\times$}          & black \\   %ok  
St.~Petersburg$^b$    &     Russia      &   40  & $BVRI$ &  347 & {\large $+$}        & orange \\   %ok    
Teide (IAC80)          &     Spain       &   80  & $VR$   &  308 & {\LARGE $\ast$}     & green \\  %ok  
Teide (LCO)            &     Spain       &   40  & $VR$   &   76 & {\large $+$}        & black \\    %ok 
Tijarafe               &     Spain       &   40  & $BR$   & 2946 & {\LARGE $\ast$}     & red \\ %ok
Vidojevica$^d$         &     Serbia      &  140  & $BVRI$ &  647 & {$\square$}         & black \\  %ok   
Vidojevica$^d$        &     Serbia      &   60  & $BVRI$ &   37 & {$\triangle$}       & black \\  %ok   
West Mountain          &     US          &   91  & $BVR$  & 2152 & {$\triangle$}       & magenta \\  %ok  
Wild Boar              &    Italy        &   24  & $VR$   &   28 & {$\triangle$}       & green \\   %ok
\hline
\end{tabular}\\
``LCO" refers to telescopes belonging to the Las Cumbres Observatory global telescope network\\
$^a$ University of Athens Observatory (UOAO)\\
$^b$ Also polarimetry\\
$^c$ Only polarimetry\\
$^d$ Astronomical Station Vidojevica\\
\end{table*}

These new data were provided as 43 different datasets (see Table \ref{tab:webt}) from 36 observatories spread in longitude around the northern hemisphere. Because the BL Lacertae observed emission is contaminated by the light of the stars in the host galaxy, the WEBT observers are invited to follow common prescriptions to perform the data reduction, in order to subtract the host contribution more easily, and get as homogeneous results as possible. These prescriptions include using an aperture radius of 8 arcsec to extract the source photometry, and an annulus of 10 and 16 arsec radii for the background. Calibration of the source magnitude is performed with respect to Stars B, C, and H of the photometric sequence by \citet{bertaud1969} in the 
%$U$ and 
$B$ band, and by \citet{fiorucci1996} in the $V$, $R$, and $I$ filters.

Because we aim to obtain accurate light curves on which a meaningful data analysis can be performed, we assembled the different datasets with extreme care, comparing the source behaviour in the four bands day by day. The density of the sampling allowed us to robustly determine whether some datasets presented offsets with respect to the bulk of the others, which can still occur notwithstanding the common reduction and calibration recipes. Such offsets were corrected by shifting the deviating datasets by their mean magnitude difference with respect to the contemporaneous data belonging to the datasets tracing the main trend. Moreover, we removed data points with large errors (more than 0.1 mag) and those that were clear outliers. The definition of an outlier is sampling-dependent, being more stringent in well-sampled nights with well-defined intranight coverage with small errors, and less stringent in less-sampled nights with more scattered data. 
%Typically, we consider a data point as an outlier if it is out by about $3 \sigma$ from the main trend. 
Finally, we reduced the data scatter by performing some binning of noisy data from the same dataset in the same night. We underline that this light curve processing is absolutely necessary if one wants to obtain a reliable dataset that can be used for further analysis.

\begin{figure}
	\includegraphics[width=\columnwidth]{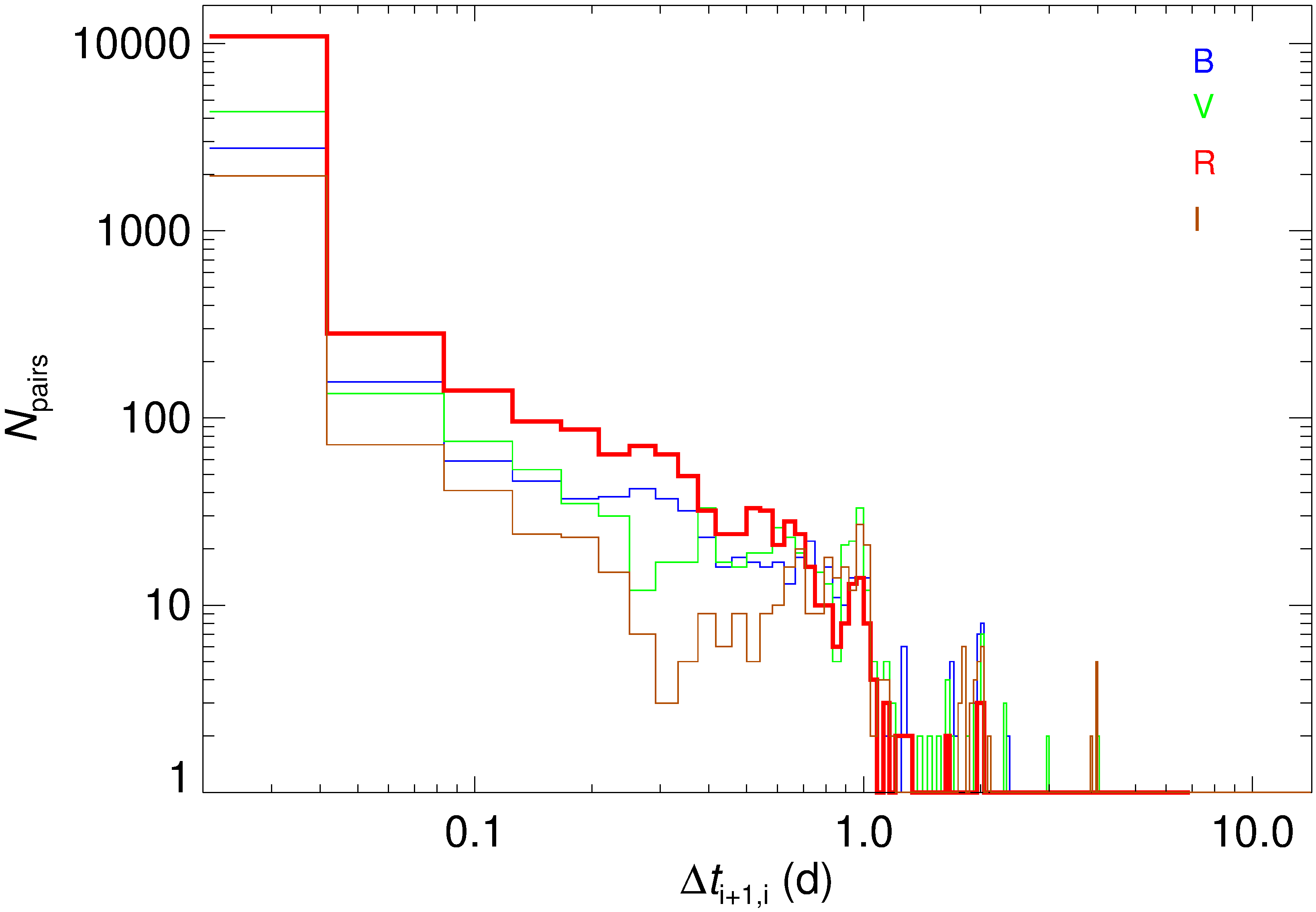}
    \caption{Distribution of the time difference between subsequent data points in the final, cleaned light curves. Blue, green, red, and brown histograms refer to the $BVRI$ bands, respectively. More than 90\% of the $R$-band data pairs have a time separation of less than 1 hour. }
    \label{fig:sampling}
\end{figure}

At the end of the procedure, we were left with 23212 data points ($\sim 94\%$ of the original sample; 3553 in $B$, 5098 in $V$, 12121 in $R$, and 2441 in $I$).
The density of data sampling can be appreciated in Fig.~\ref{fig:sampling}, where the time difference between subsequent data points is plotted for all bands in bins of 1 hour each. In the case of the best-sampled $R$ bands, more than 90\% of data pairs are contained in the first bin.

\begin{figure*}
 \includegraphics[width=12cm]{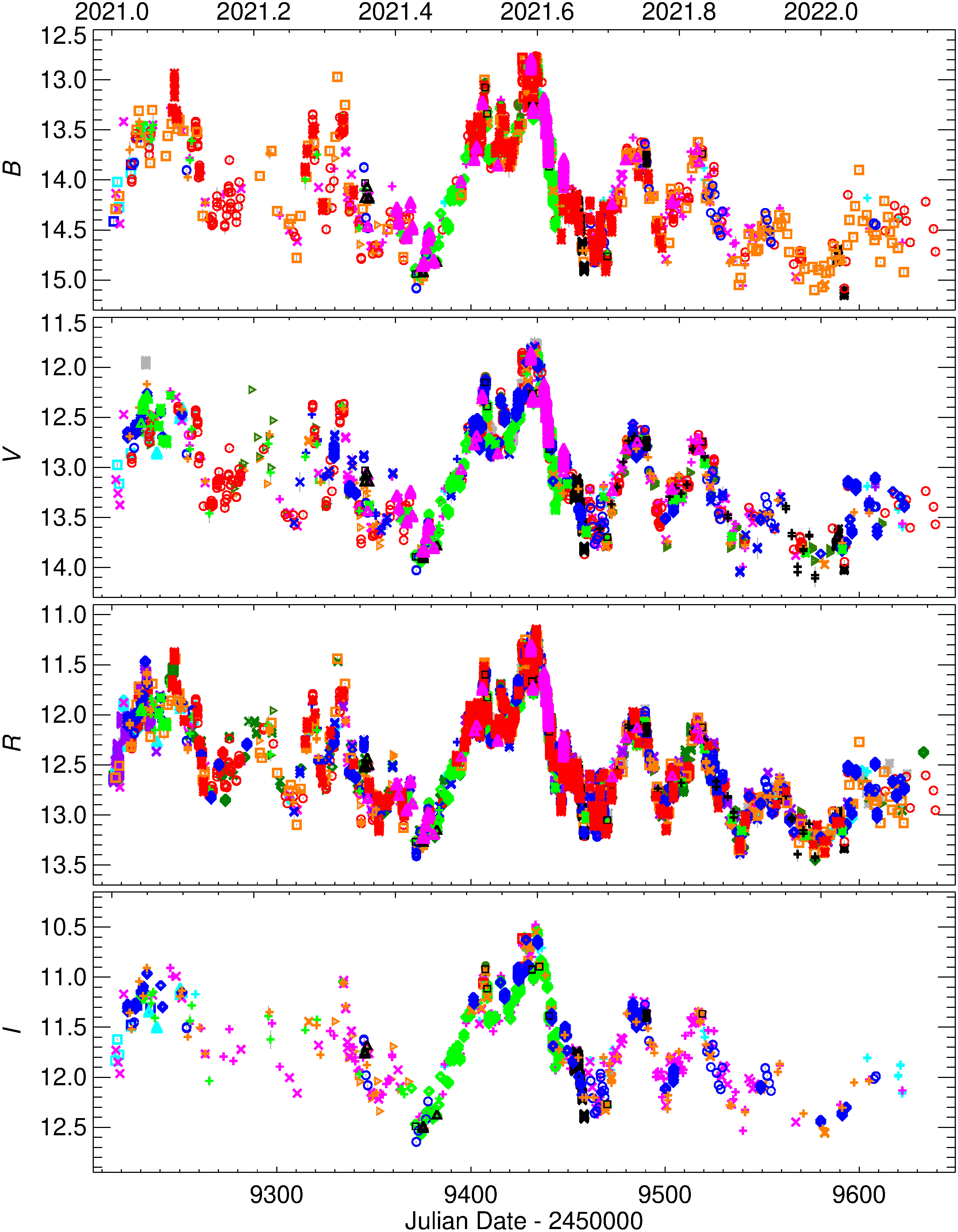}
 \caption{WEBT multiband optical light curves (observed magnitudes) of BL Lacertae during the 2021--2022 observing season. Different colours and symbols are used to distinguish the contributing datasets, as specified in Table \ref{tab:webt}. Measurement uncertainties are plotted in grey and are usually smaller than the datapoint size.}
 \label{fig:webt_2021}
\end{figure*}

\begin{figure*}
 \includegraphics[width=12cm]{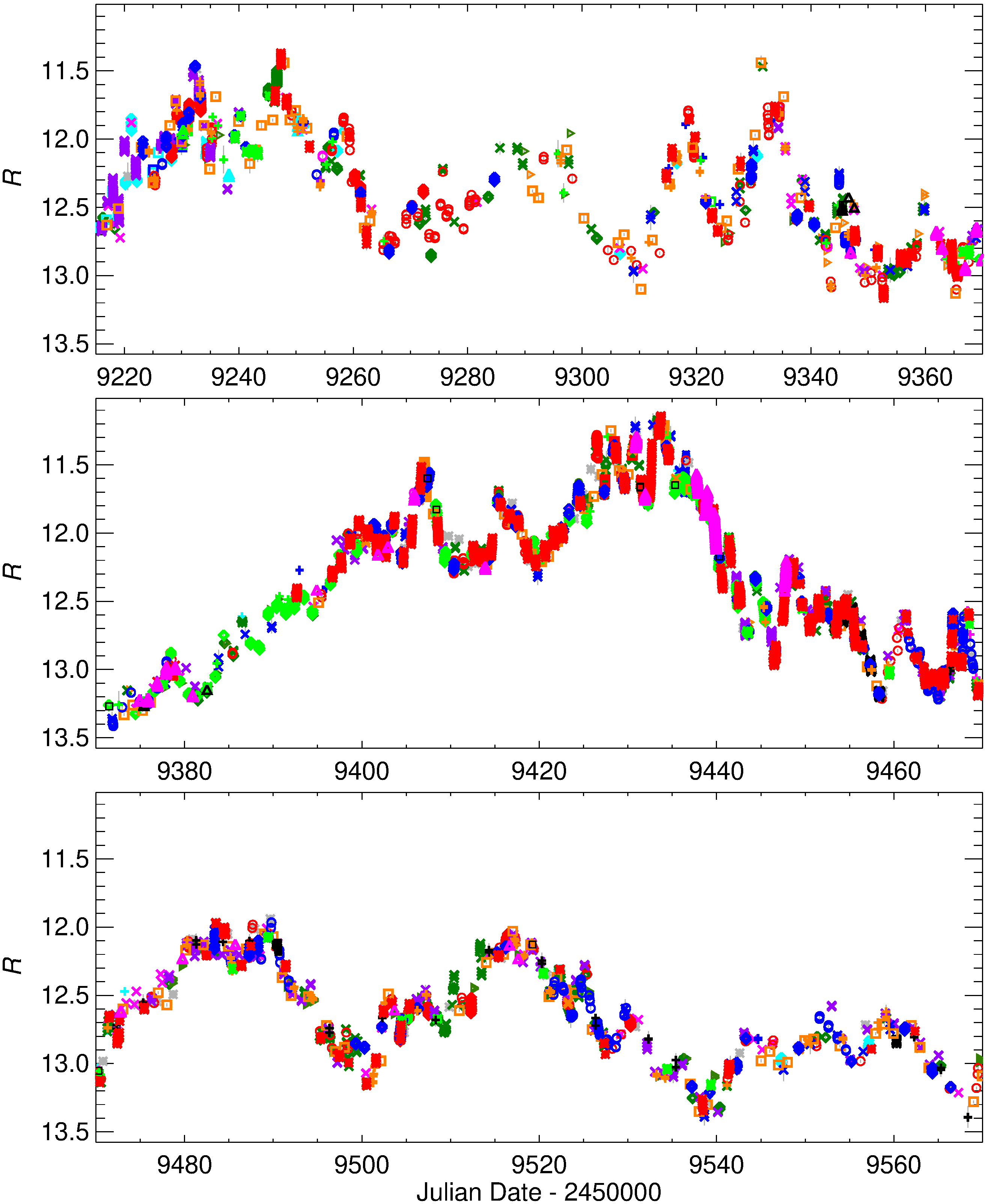}
 \caption{An enlargement of the $R$-band light curve shown in Fig.~\ref{fig:webt_2021}.}
 \label{fig:outburst}
\end{figure*}

The resulting cleaned light curves are shown in Fig.\ \ref{fig:webt_2021} in observed magnitudes.
The peak of the mid 2021 outburst represents the observed historical maximum, 
%at least from 1968, %\citep[see][]{villata2004a}, 
with $B=12.75 \pm 0.02$, $V=11.75 \pm 0.02$, $R=11.14 \pm 0.03$, and $I=10.47 \pm 0.01$ on August 6-7.
The overall variation amplitude, defined as the maximum minus minimum magnitude, is 2.40, 2.36, 2.31, and 2.17 in the $BVRI$ filters, respectively. Although the light curves have different sampling, the increasing amplitude variation with increasing frequency is typical of BL Lac objects.
An enlargement of the $R$-band light curve of Fig.~\ref{fig:webt_2021} is shown in Fig.~\ref{fig:outburst} to better appreciate the source short-term variability.

In the following analysis we will also use flux densities in mJy. They were obtained from the observed magnitudes using the zero-mag flux densities by \citet{bessell1998}, after correcting for the Galactic extinction values given by the NASA/IPAC Extragalactic Database\footnote{https://ned.ipac.caltech.edu/} (NED): 1.192, 0.901, 0.713, and 0.495 mag in the $BVRI$ bands, respectively. We also corrected for the emission contribution of the host galaxy according to \citet{raiteri2009,raiteri2010}, i.e.\ we adopted a host galaxy flux density of 1.297, 2.888, 4.229, 5.903 mJy in the $BVRI$ bands, respectively, and subtracted 60\% of this flux, because this is the amount of contamination for an aperture radius of 8 arcsec with background taken in an annulus of 10 and 16 arcsec radii, as recommended by the WEBT prescriptions.

\section{Intraday variability}
\label{sec:idv}
%Vedi notti (9401) 9407 9432 (9436) 9437 9438 9439 (9447)
%perkins 9466 9469
%9454 9455 9456 9457 9458 9490 dcf b-r e bmr su galsub
BL Lacertae is one of the blazars which are known for their noticeable short-term variability and, in particular, for their intraday variability (IDV). Many IDV episodes have been reported in the WEBT studies on this source cited in the Introduction and in other works \citep[e.g.][]{miller1989,meng2017,fang2022,imazawa2022}.

\begin{figure}
	\includegraphics[width=\columnwidth]{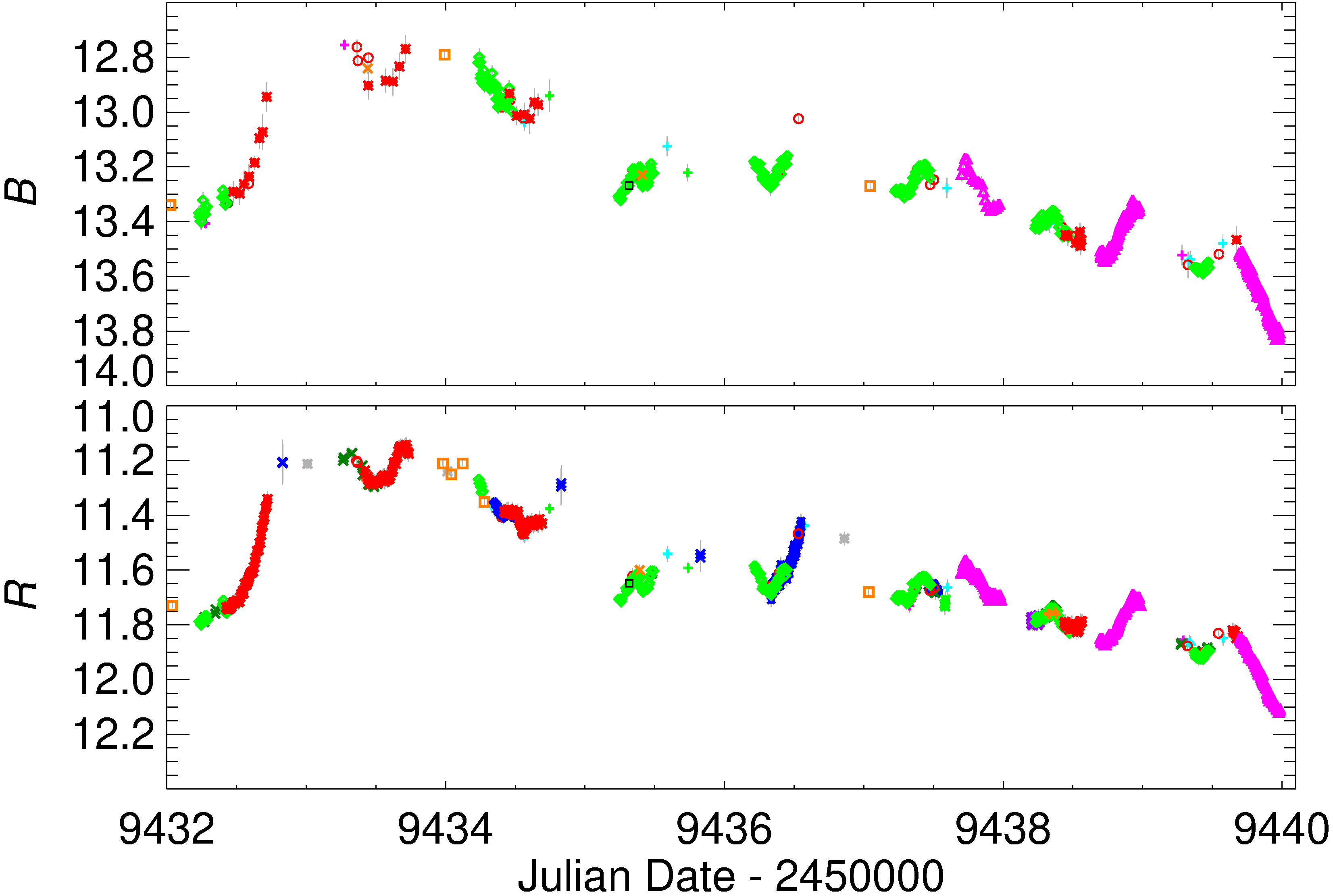}
    \caption{A detailed view of a portion of the $B$ (top) and $R$ (bottom) light curves covering the time period 2021 August 5 to 13, in which many IDV episodes can be seen thanks to the high-time resolution sampling. Measurement uncertainties are plotted in grey and are usually smaller than the datapoint size.}
    \label{fig:br_32-40}
\end{figure}

\begin{figure}
	\includegraphics[width=\columnwidth]{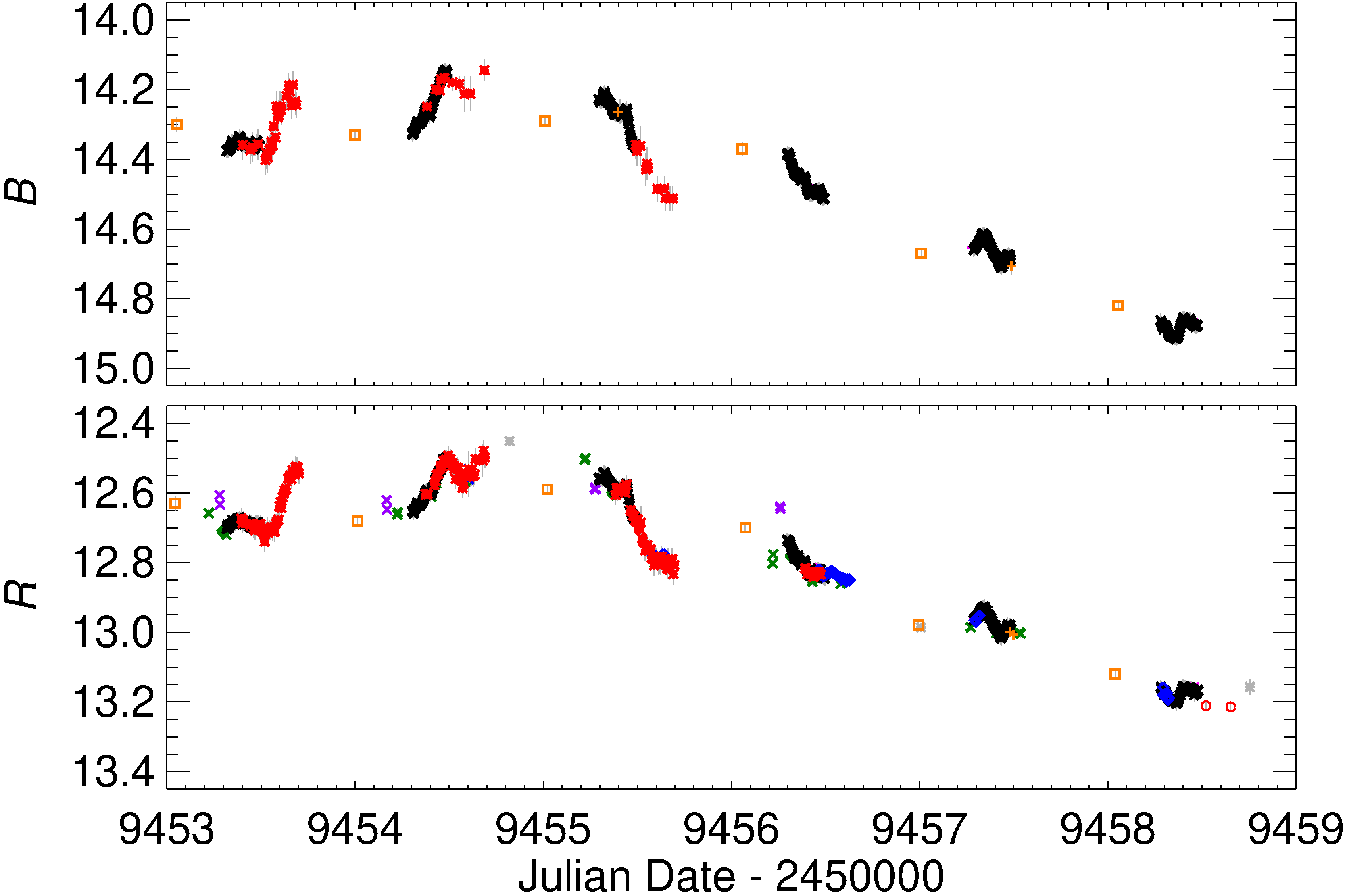}
    \caption{As Fig.~\ref{fig:br_32-40}, but for the period 2021 August 26 to September 1.}
    \label{fig:br_53-59}
\end{figure}

\begin{figure}
	\includegraphics[width=\columnwidth]{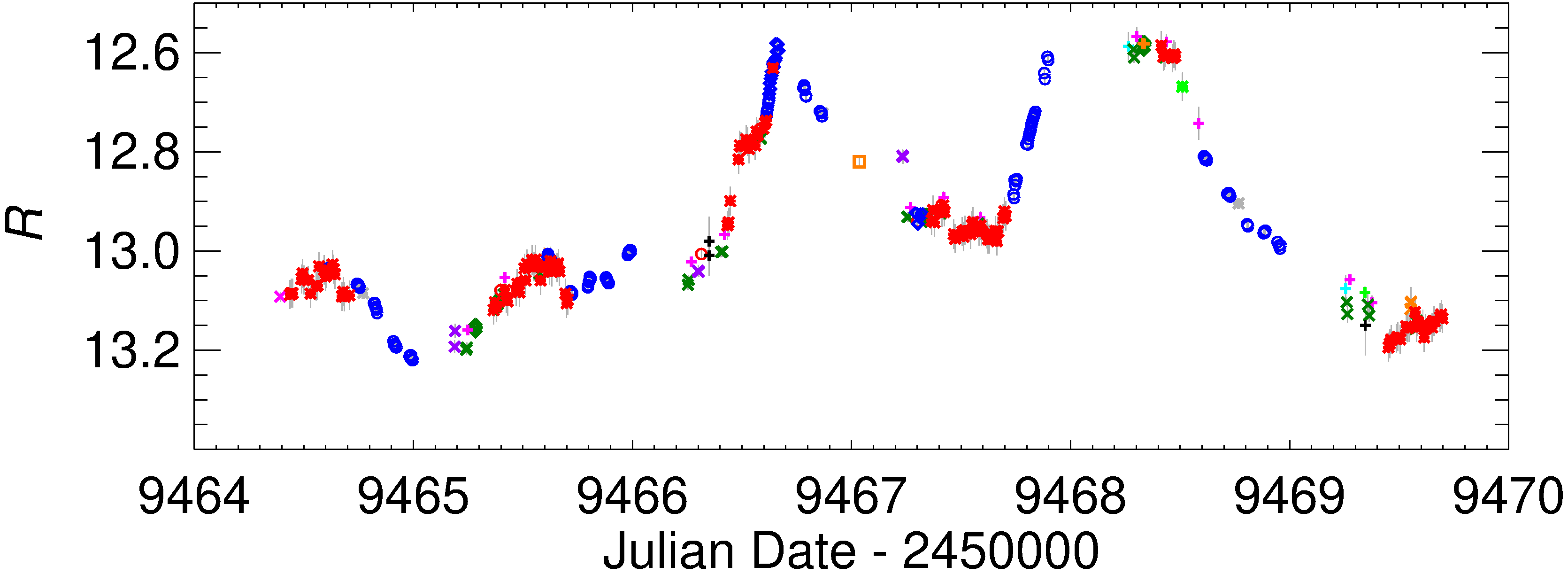}
    \caption{A noticeable example of IDV in the $R$ band in the period 2021 September 6 to 12.}
    \label{fig:r_64-70}
\end{figure}

In the period considered in this paper, further episodes of IDV were observed, particularly during the brightest phases. Interesting examples are shown in Fig.~\ref{fig:br_32-40} and Fig.~\ref{fig:br_53-59}, where not only the $R$-band, but also the $B$-band light curve is very well sampled. An additional example of the very closely spaced time-sampling obtained by the WEBT Collaboration in the $R$ band is shown in Fig.~\ref{fig:r_64-70}.
Here it is particularly evident that the participation of observers well distributed in longitude can lead to an almost continuous monitoring of the source, revealing the details of its short-term variability.

\begin{figure}
	\includegraphics[width=\columnwidth]{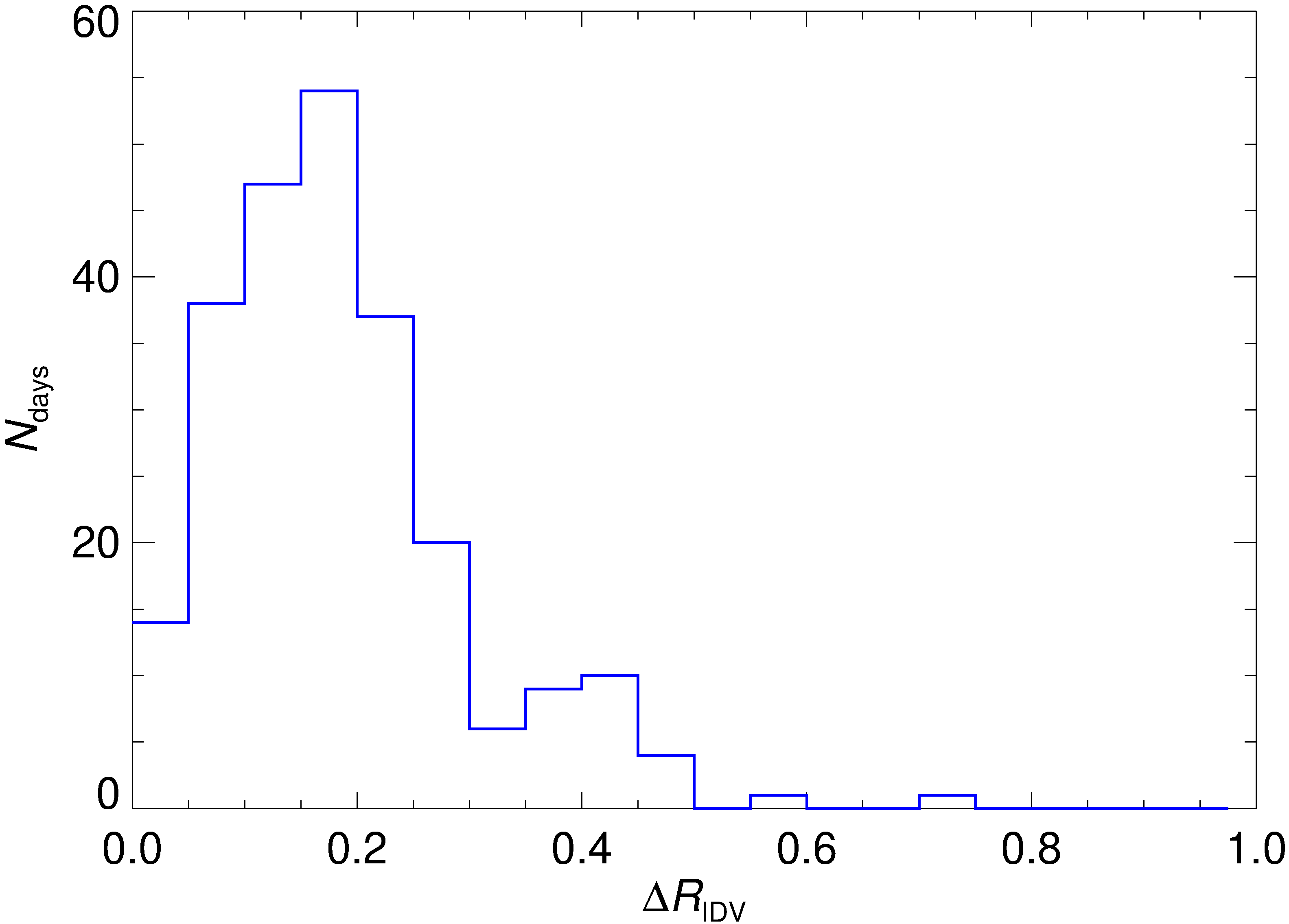}
    \caption{Number of days where the IDV magnitude amplitude in the $R$ band  is $\Delta R_{\rm IDV}$, considering the 241 days where more than ten observations were obtained in the same day.}
    \label{fig:idv_stat2}
\end{figure}

In Fig.~\ref{fig:idv_stat2}, we show the number of days for which an observed maximum minus minimum $R$-band magnitude difference, $\Delta R_{\rm IDV}$, was seen for all days having a number of observations greater than ten.
 %Because of the distribution in longitude of the WEBT observer, a day here is defined as the interval between $t_i-0.1$ and $t_i+0.9$, where $t_i$ indicates the days in the light curve at 0 UT. 
In 73\% of the 241 resulting cases, the IDV amplitude ranges from 0.05 to 0.25 mag, but there are 51 days in which it is greater than 0.25 mag, with a maximum value of $\sim 0.7$ mag.
These are of course lower limits, because they refer to what was observed, the actual total variability amplitude likely being larger.

\begin{figure}
	\includegraphics[width=\columnwidth]{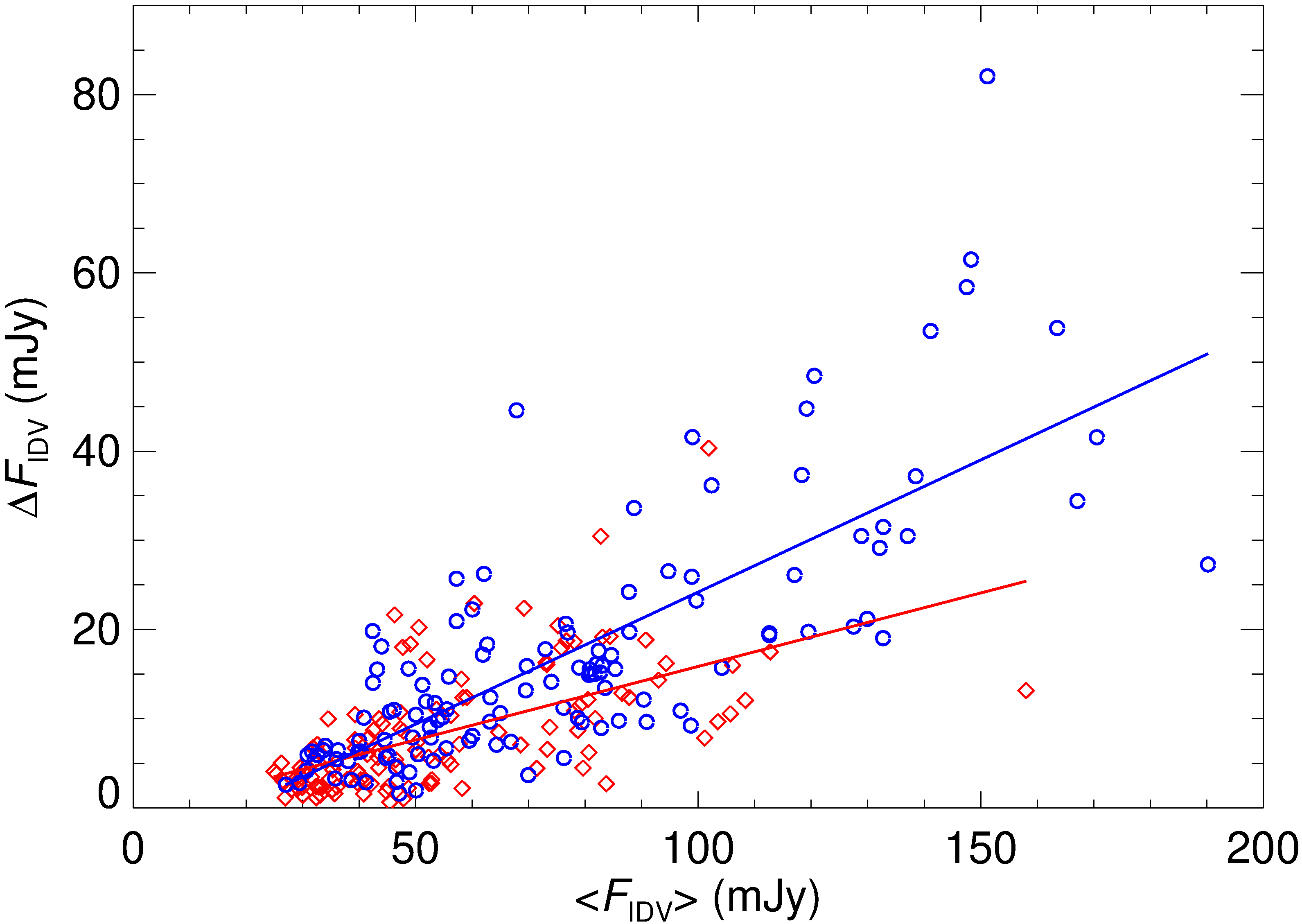}
    \caption{IDV flux density amplitude in the $R$ band as a function of the mean IDV flux density for the 241 days where more than ten observations were obtained. Colours distinguish days with $<30$ data points (red diamonds) from those with $\ge 30$ observations (blue circles). Lines represent linear fits to the two samples.}
    \label{fig:idv_stat3}
\end{figure}

When considering the IDV flux amplitude in the $R$ band, $\mathbf{\Delta F_{\rm IDV}}$, as a function of the average flux density in the same band, we obtain the result plotted in Fig.~\ref{fig:idv_stat3}, which suggests that the IDV amplitude increases with brightness. In the figure, we distinguish the cases where the number of intraday observations is $<30$ from those where it is $\ge 30$, and plot the corresponding linear fits to highlight that on average the amplitude increases with sampling.
The different slope between the two fits stresses how the observed IDV amplitudes must be considered as lower limits to the total source IDV variations.

We note that larger flux variations in the brightest states is what one would expect if the long-term variability were due to changes of the Doppler factor \citep{raiteri2017_nature}. This will be further discussed in the following sections.

\section{Spectral variability}
\label{sec:spec}

Blazars are known to exhibit specific spectral trends. 
In particular, BL Lac objects usually show a bluer-when-brighter behaviour.
In the case of BL Lacertae, already \citet{villata2002} and \citet{villata2004a} recognised that the long-term variability is almost achromatic, while the short-term variability shows a chromatic behaviour. Similar results have been found for other blazars, including S5~0716+714 \citep{raiteri2021a} and S4~0954+658 \citep{raiteri2021b}.

\begin{figure}
	\includegraphics[width=\columnwidth]{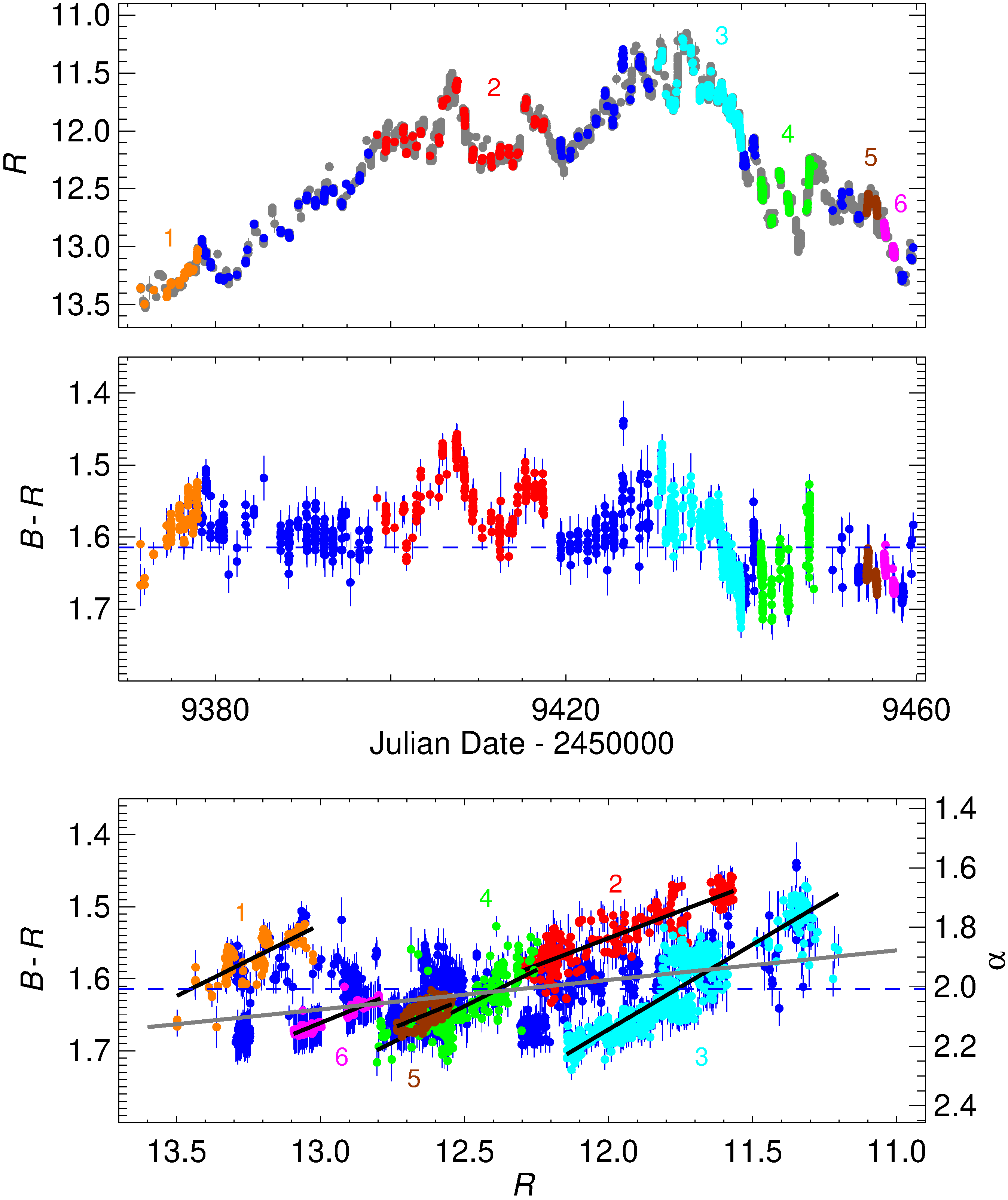}
    \caption{Top: $R$-band magnitudes during the main mid-2021 outburst, with the host galaxy emission contribution subtracted. Middle: $B-R$ colour indices versus time. Bottom: the same colour indices versus brightness. Periods of strong chromatic variability are labelled by number and highlighted with different colours. In the bottom panel the black lines represent linear fits to the coloured data points, while the grey line refers to all the data shown in the figure.}
    \label{fig:bmr}
\end{figure}

Colour indices were obtained by first correcting the light curves for the host galaxy emission contribution and then coupling data points with errors smaller than 0.03 mag in the two filters, which were taken by the same telescope within 10 min. In this way we obtained 2629 $B-R$ colour indices, with values ranging from 1.42 to 1.79 and average $<B-R> = 1.61$ with standard deviation of 0.05 mag. 
In Fig.~\ref{fig:bmr} we show the $B-R$ colour indices plotted against both time and brightness, for the most well-sampled period corresponding to the major mid-2021 outburst.  

In the last panel of Fig.~\ref{fig:bmr}, we also show the optical spectral index $\alpha$ (assuming $F_\nu \propto \nu^{-\alpha}$) derived from the colour index. Its values range from 1.53 to 2.43, and the mean value is $<\alpha>=2.01$ with a standard deviation of 0.13. This implies that the optical SED is steep/soft, and therefore the peak of the synchrotron emission bump is located at lower frequencies, as expected in a low-energy peaked BL Lac object \citep[see][]{raiteri2009,larionov2010,raiteri2010,raiteri2013}.

Furthermore, we identified examples of short-time intervals characterised by strong spectral variations; they are highlighted in Fig.~\ref{fig:bmr}.
%Furthermore, in Fig.~\ref{fig:bmr}, we highlighted data points belonging to time intervals that are characterised by short-term flux and colour variability. 
In these time intervals the source shows a clear bluer-when-brighter trend, with linear fit slopes ranging from 0.15 to 0.24. In contrast, a linear fit to the whole sample in the considered period indicates an almost achromatic behaviour, with a slope of only 0.04.

According to the interpretation given in the papers cited at the beginning of this section, the achromatic long-term trend would be due to Doppler factor variations of geometrical origin, while the chromatic short-term variability would be produced by intrinsic energetic processes.

\section{Variability time scales}
\label{sec:wavelet}
Wavelet analysis is a powerful tool for detecting variations of power in time series, especially when they are not persistent, but occur only during defined time spans.
The method is commonly adopted to unveil possible periodicities in blazar light curves at different frequencies and on a variety of time scales \citep[e.g.][]{gupta2009,zhou2018,gupta2019,penil2020,otero2020,jorstad2022,roy2022,otero2022}.

\begin{figure*}
 \includegraphics[width=15cm]{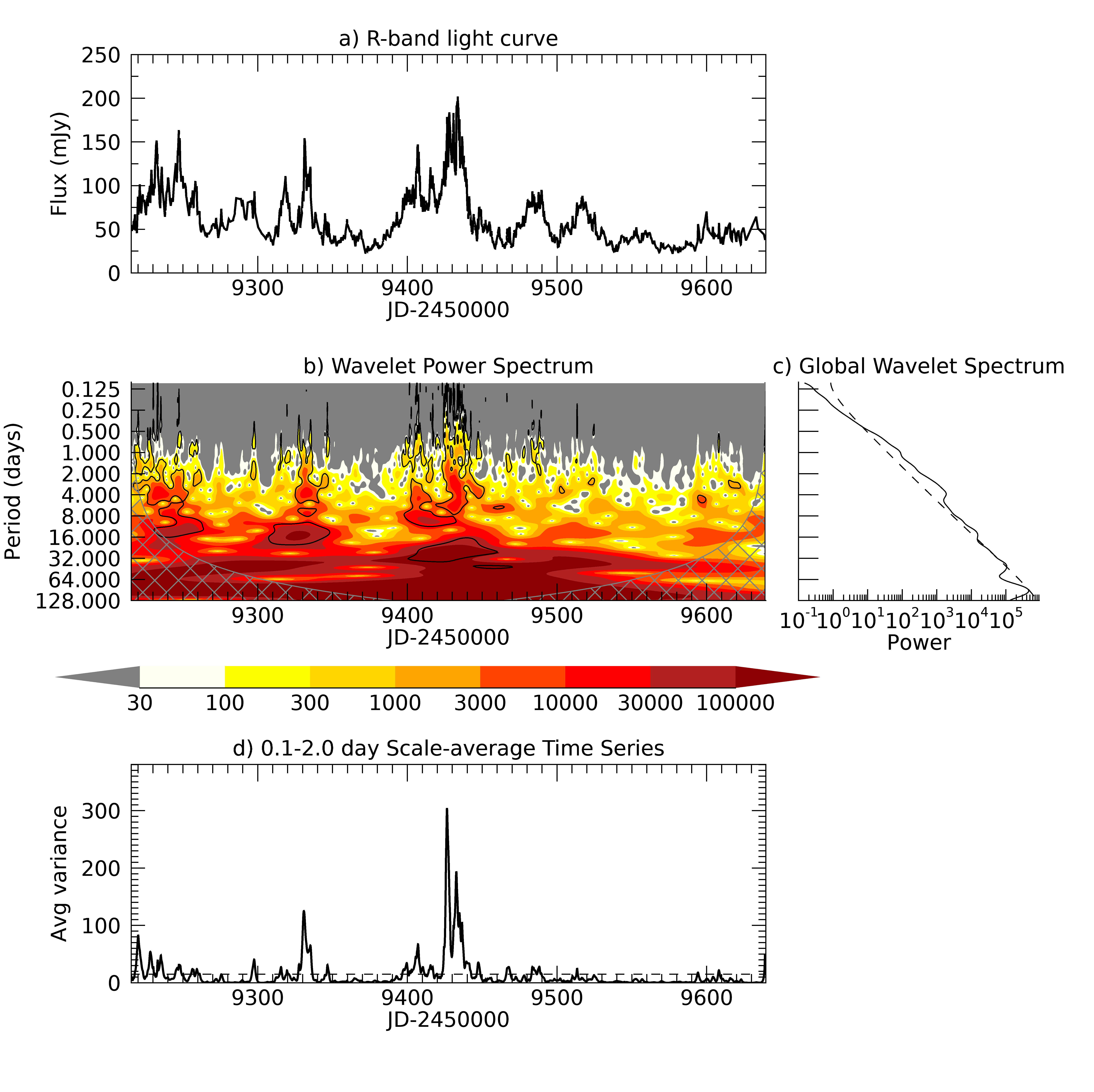}
 \caption{Results of the wavelet analysis. a) $R$-band flux densities of BL Lacertae in the 2021-2022 observing season. b) Wavelet power spectrum: the strength of the power is colour-coded according to the underlying palette; black contours define regions with confidence greater than 99\%; the grey grid represents the ``cone of influence" affected by edge effects. c) Global wavelet spectrum with 99\% confidence level marked by a dashed line. d) Time series of averaged periods between 0.1 and 2 days, with 99\% confidence level represented by the dashed line.}
 \label{fig:wavelet}
\end{figure*}

To investigate the presence of characteristic variability time scales in the BL Lacertae optical emission, we performed a wavelet analysis on the $R$-band flux density light curve. The results are shown in Fig.~\ref{fig:wavelet}; they are based on the algorithm implementation by \citet{torrence1998}. The wavelet power spectrum shows a series of significant periods, with confidence greater than 99\%, ranging from about one month down to a few hours. These periodic signals appear in correspondence of the three major flaring states, when the flux density exceeded about 100 mJy. 
In particular, the shortest scales (0.1-2 days) are more evident when the flux is higher.
This is also highlighted by the plot of the variance, where the power is averaged over periods between 0.1 and 2 days. Indeed, the variance largely exceeds the 99\% confidence level in the brightest states.
The global wavelet spectrum, which represents an average of the power in time over the whole period, shows that time scales from about half a day to 4 days are particularly strong, i.e.~well above the 99\% confidence level.
%In summary, short time scale variability appears only during the brightest states of the source. 
%This is expected if outbursts are produced by a better alignment of the optical emitting region, which implies a larger Doppler factor, and thus a decrease of the variability time scales, since $\Delta t = \Delta t' / \delta$.
These results are in agreement with those of the wavelet analysis performed by \citet{jorstad2022}, who found short-term QPOs of 0.55, 4, and 14 days during the 2020 outburst, and indicate that fast quasi-periodic flux oscillations are common in major flaring states of BL Lacertae.
%We note that the QPOs with period of 0.55 d detected by \citet{jorstad2022} in the light curve of BL Lacertae in 2020 were found during the first stage of a major outburst.

\section{Optical polarimetry}
\label{sec:pola}

Interpreting the polarisation variability in blazars has always been a challenge. Sometimes the optical flux and polarisation degree, $P$, are observed to be correlated, sometimes they seem anticorrelated, and often they are uncorrelated \citep[e.g.][]{raiteri2012,raiteri2013,raiput2022}.
Several studies concentrated on the wide rotations of the electric vector position angle \citep[EVPA, see e.g.][]{larionov2008,marscher2008,marscher2010,sasada2012,raiteri2013,blinov2015,carnerero2015,larionov2016,raiteri2017,larionov2020} that in some occasions appear correlated with $\gamma$-ray flares \citep{blinov2018} and may be produced by magnetic reconnection \citep{zhang2018}.
But some large rotations of EVPA can also be the result of stochastic processes, i.e.~turbulence \citep[e.g.][]{marscher2014,blinov2015,kiehlmann2016,kiehlmann2017,raiteri2017}.

During the 2021--2022 observing season, 1075
optical polarimetric data of BL Lacertae were acquired in the $R$ band at the 60~cm telescope of the Belogradchik Observatory, at the 70~cm and 260~cm telescopes of the Crimean Observatory, at the 256~cm Nordic Optical Telescope (NOT), at the 180~cm Perkins telescope, and at the 40~cm telescope of the St.~Petersburg Observatory (see Table~\ref{tab:webt}).

\begin{figure*}
 \includegraphics[width=12cm]{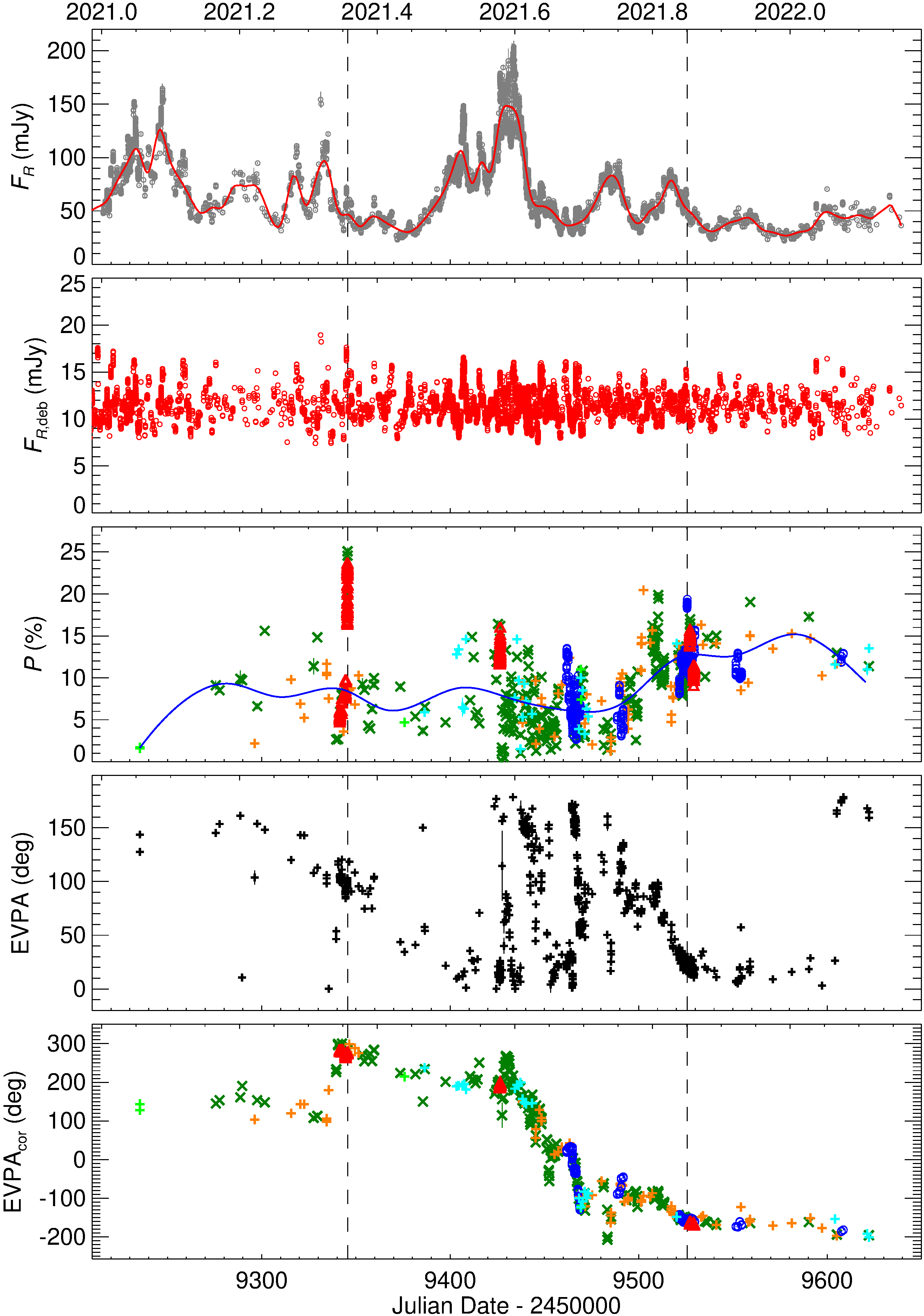}
 \caption{From top to bottom: i) Optical flux densities in the $R$ band; the red line represents the cubic spline interpolation on the binned light curve, with variable binning depending on brightness. ii) Optical flux densities after deboosting as explained in Sect.~\ref{sec:snake}. iii) Observed optical polarisation degree; the blue line represents a cubic spline interpolation through the 30-day binned data to highlight the long-term trend. iv) EVPA constrained between 0\degr\ and 180\degr. v) EVPA after arrangement for the $\pm n \times 180\degr$ ambiguity. In the third and last panels, different colours and symbols are used to distinguish the contributing datasets, as specified in Table \ref{tab:webt}. The vertical dashed lines mark the two events that are discussed in the text.}
 \label{fig:pola}
\end{figure*}

In Fig.~\ref{fig:pola}, $P$ and the EVPA are shown as a function of time, and are compared to the behaviour of the optical flux density
%(both before and after deboosting) 
in the same band. 
%Fig.~\ref{fig:pola_zoom} displays an enlargement of the better-sampled period to better distinguish the main features.
The observed $P$ ranges from 0.025\% to 25\%, with a mean value of 10\% and a standard deviation of $\sim 4\%$.
The dilution effect of the unpolarised light of the host galaxy \citep{raiteri2021_galaxies} has a negligible effect because of the source brightness.

The EVPA is known to present a $\pm n \times 180\degr$ ambiguity, which makes the reconstruction of its behaviour challenging, if the sampling presents interruptions. In the case of blazars, it has been shown that the sampling should be at least daily to obtain robust results \citep{kiehlmann2021}.
To handle this problem, we first constrained all EVPA values between $0\degr$ and $180\degr$.
%by adding / subtracting $180\degr$. 
Then, when the absolute value of the difference between the value of EVPA at a given time and its predecessor exceeded $90 \degr$, we added/subtracted $180 \degr$ to minimise this difference.
%stesso risultato se nella differenza consideriamo gli errori
We note the following interesting features: 
\begin{itemize}
    \item The degree of polarisation $P$ shows considerable IDV.
    \item On $\rm JD=2459345.55$, $P$ rapidly reaches the maximum value seen during our 2021-2022 observations. This does not correspond to a major event in flux, but rather to a minor peak.
%    \item Other $P$ maxima have minor peak counterparts in $F_R$.
    \item Starting from $\rm JD \sim 2459500$, $P$ shows an increasing trend of the baseline level, which is opposite to the trend of the optical flux density. 
    \item The EVPA displays noticeable variability.
    \item When the $\pm n \times 180\degr$ ambiguity is fixed as explained above, an extremely wide rotation of the EVPA ($\ga 360 \degr$) appears, starting from about the middle of the major 2021 outburst, i.e.~$\rm JD \sim 2459430$. This EVPA rotation proceeds in steps.

\end{itemize}

We note that the anticorrelation between the bulk behaviour of $P$ and $F_R$ was explained by \citet{raiteri2013} in the framework of a geometrical scenario (where long-term flux variations are produced by changes of the viewing angle), by assuming a low value of the Lorentz factor. They also showed that the same geometric interpretation, but with a high Lorentz factor, can instead lead to correlation between flux and polarisation degree.

\begin{figure}
	\includegraphics[width=\columnwidth]{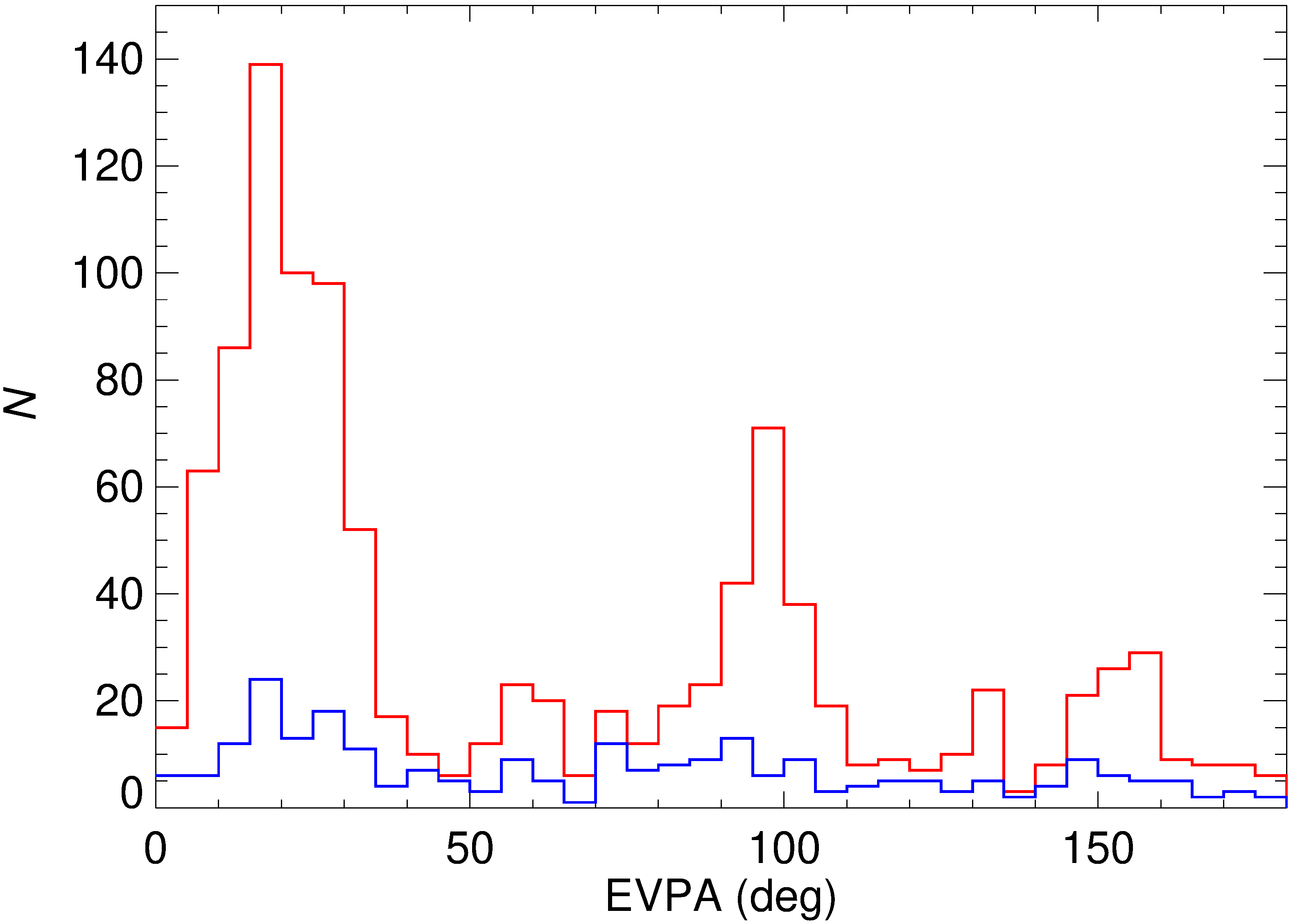}
    \caption{Distribution of EVPA values (red line). The major peak at 15-20\degr\ corresponds to the preferential direction found by \citet{raiteri2013} and \citet{jorstad2022}. The blue line shows the distribution of the mean EVPA values averaged over 6-h bins to check for the effect of sampling. Binning strongly reduces the dominance of the preferential direction and makes the distribution of EVPA values more uniform overall.}
    \label{fig:stat_evpa}
\end{figure}

We also note that BL Lac sources generally show a preferred EVPA direction \citep{smith1996}, and in BL Lacertae this was found to be 15-25\degr\  \citep{blinov2009,raiteri2013,jorstad2022}, indicating a magnetic field roughly perpendicular to the direction of the radio jet at 43 GHz \citep{jorstad2022}. In the period of strong activity that we are considering, the EVPA shows intense variability. If we plot the distribution of the EVPA values (see Fig.~\ref{fig:stat_evpa}), a concentration around 15-20\degr\ appears, so it seems that the above preferred direction still dominates. However, this is mostly due to the sampling. Indeed, if we average the EVPA values over 6-h bins, the importance of the 15-20\degr\ peak is strongly reduced and the distribution of EVPA values appears more uniform overall.
%This is better seen in Fig.~\ref{fig:stat_evpa}, where the concentration around 15-20\degr\ is evident, while the peak at 90-105\degr\ is mainly due to the 48 data points acquired on $\rm JD=2459345$.

We finally note that large jumpy rotations of the radio EVPA of 0727-115 were reported by \citet{aller1981}, who suggested a rotation or quasi-circular motion in the source emitting
region. Moreover, a wide, step-like rotation of the optical EVPA observed in the BL Lac object S4 0716+714 was successfully reproduced by \citet{larionov2013} assuming a shock wave propagating in a helical jet or along a helical path in the jet.

\section{The twisting jet}
\label{sec:snake}
In previous papers, we have proposed that long-term variability in blazars has a geometrical origin, due to changes in the Doppler factor produced by variations in the orientation of the jet with respect to the line of sight \citep[e.g][]{villata2002,villata2007,raiteri2013,raiteri2017_nature,raiteri2021b}. 
Indeed, the flux density at a given frequency $\nu$ depends on the Doppler factor according to $F_\nu \propto \delta^{n+\alpha}$,
where $n=2$ for a continuous jet \citep{urry1995} and $\alpha$ is the optical spectral index (see Sect.~\ref{sec:spec}). As mentioned in the Introduction, the Doppler factor $\delta$ depends on the viewing angle, so that variations in the jet orientation translate into changes in the source flux. 
Our model envisages an inhomogeneous and twisting jet. 
Inhomogeneous means that synchrotron radiation of increasing wavelength is emitted from jet regions that are at growing distance from the supermassive black hole, due to the interplay between emitted and absorbed frequencies, with opacity decreasing downstream the jet. This is suggested by the increasing time delay with which the radio flux variations at increasing wavelength follow those observed in the optical band in correlated optical-radio flares. 
Moreover, according to the model, various emitting regions have different viewing angles, which implies that the jet is curved, possibly helical. This comes from the fact that at a given epoch the brightness state can vary a lot with frequency, which is ascribed to different Doppler boosting. Finally, the orientation of the emitting regions changes over time, leading to a twisting jet.
This scenario can be figured out by considering a rotating helical jet, where the minimum-viewing-angle (maximum Doppler boosting) zone shifts along the jet as the helix rotates.
Starting, for example, with a minimum viewing angle in the optical emitting region, we will observe an outburst in the optical band. But then the helix rotates, and the minimum viewing angle is progressively achieved by regions emitting longer and longer wavelength radiation, so that we will see time-delayed correlated events at these wavelengths.
The most notable application of this model was the interpretation of the long-term variability behaviour of CTA~102 by \citet{raiteri2017_nature}.

It is worth mentioning that there may be several reasons why the jet is twisting, including orbital motion in a binary black hole system, jet precession, or plasma instabilities inside the jet. There is much evidence presented in many papers which support this interpretation, both observational \citep[e.g.][]{perucho2012,fromm2013,casadio2015,britzen2017,britzen2018,issaoun2022,zhao2022} and theoretical \citep[e.g.][]{begelman1980,nakamura2001,hardee2003,moll2008,mignone2010,liska2018}.

In this paper we found hints in favour of the above geometrical scenario for the long-term variability of BL Lacertae in 2021-2022.
In Sect.~\ref{sec:idv} we saw that the variability amplitude increases with flux, and this is in agreement with the prediction that $\Delta F_\nu \propto \delta^{n+\alpha}$.
Moreover, if the long-term trend is due to geometrical reasons, then the long-term spectral variability must be almost achromatic, as found in Sect.~\ref{sec:spec}.
Also the anticorrelation between the long-term trends of the flux and degree of polarisation (Sect.~\ref{sec:pola} and \ref{sec:corre}) fits well in a geometric scenario, provided a relatively low Lorentz factor is assumed, as seems adequate for BL Lacertae \citep{jorstad2005,raiteri2013}. 
%And this is indeed the case for BL Lacertae \citep{jorstad2005}
Finally, we found the appearance of short variability time scales only during the source brightest states, which can be explained by the decrease in the time scales due to Doppler beaming $\Delta t = \Delta t' / \delta$, where the prime refers to the rest frame.
In conclusion, all these properties can be explained by assuming that the long-term behaviour of BL Lacertae is produced by variations of the Doppler factor of the jet-emitting regions due to changes in their orientation.

%If we then assume that the long-term trend is due to Doppler factor variations, this affects not only the fluxes but also the time scales. A larger Doppler factor implies higher fluxes, larger variability amplitudes, and shorter time scales.For a continuous jet \citep{urry1995}:$F_\nu (\nu)=\delta^{2+\alpha} \, F'_{\nu '} (\nu)$,where primed quantities refer to the rest frame and $\alpha \sim 2$ is the spectral index (see Sect.~\ref{sec:spec}).
%Moreover, the variability amplitude of intrinsic flux changes increases as $\Delta F \propto \delta^{2+\alpha}$, and indeed we saw in Sect.~\ref{sec:idv} that IDV is larger in brighter states.
%Finally, $\Delta t = \Delta t' / \delta$, and in Sect.~\ref{sec:wavelet} we found the appearance of short-term time scales only during the source brightest states.

Therefore, we inferred the amount of variable beaming from the observed light curve, following the method explained in \citet{raiteri2017_nature}. 
Because relativistic beaming produces both a flux increase and a decrease of time scales, we applied a variable binning to the $R$-band flux density light curve, where the time bin is reduced depending on the source brightness. Then we used a cubic spline interpolation through the binned light curve to obtain a model for the long-term behaviour, which we assume to reflect the variations of the Doppler factor.
This was then used to correct the flux densities for the effect of the relativistic beaming.
The deboosted flux densities are plotted in Fig.~\ref{fig:pola} and show an almost constant variability amplitude. 
This further supports the validity of the Doppler changing scenario.
We ascribe this residual short-term variability to physical energetic processes occurring inside the jet.

\begin{figure}
	\includegraphics[width=\columnwidth]{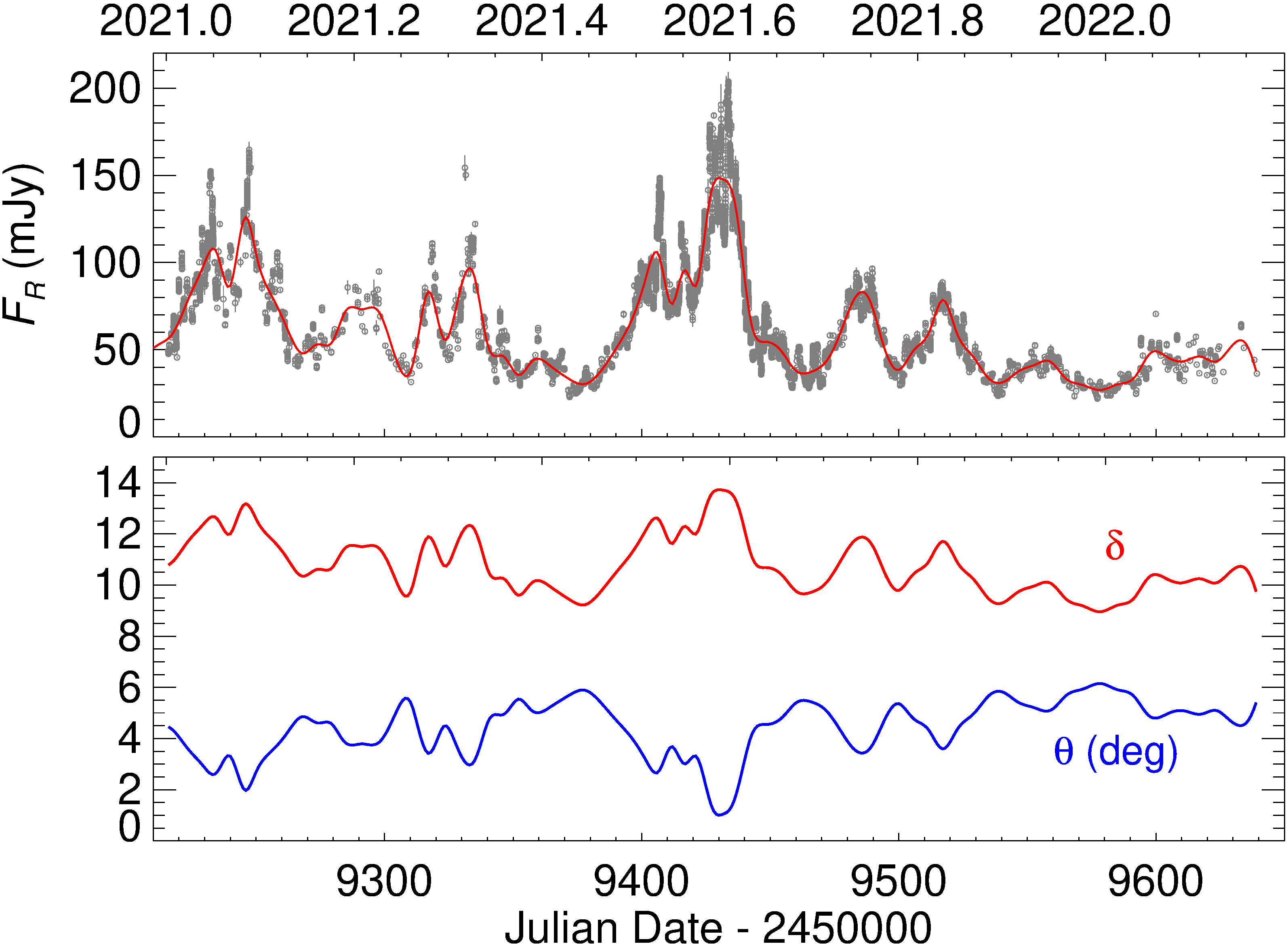}
    \caption{Top: Optical flux densities in the $R$ band; as in Fig.~\ref{fig:pola}, the red line represents the cubic spline interpolation on the binned light curve, with variable binning depending on brightness, and is assumed to represent the long-term trend. Bottom: Behaviour of the Doppler factor $\delta$ and the viewing angle $\theta$ in time derived from the long-term trend.}
    \label{fig:delta_teta}
\end{figure}

We finally note that in order to infer the behaviour of $\delta(t)$ and $\theta(t)$ from the long-term trend, we must assume a value for both the Lorentz factor and, for example, the minimum viewing angle. We adopted $\Gamma=7$ \citep{jorstad2005,raiteri2013} and $\theta_{\rm min}=1 \degr$, which implies $\delta_{\rm max}\approx 13.73$. Then
$\delta(t)=\delta_{\rm max} \,(F_{\rm spline}(t)/F_{\rm spline,max})^{1/(2+\alpha)}$, and 
$\theta(t)=\arccos[(\Gamma \, \delta(t)-1)/(\delta(t) \sqrt{\Gamma^2-1})]$.
The resulting $\delta(t)$ and $\theta(t)$ are shown in Fig.~\ref{fig:delta_teta}.

\section{Exploring the flux-polarisation correlations}
\label{sec:corre}
In Sect.~\ref{sec:pola} we noticed a general anticorrelation between the overall behaviour of the flux density and degree of polarisation. We now want to investigate their relationship in more detail.
%As a first step, we analyse the correlation between the $R$-band flux densities and the polarisation degree shown in Fig.~\ref{fig:pola}.
To this end, we use the discrete correlation function \citep[DCF,][]{edelson1988,hufnagel1992}, which is a method specifically conceived for unevenly-sampled data trains.
Our implementation of the algorithm performs a local normalisation of the DCF, for more accurate results \citep{peterson2001,max2014}.

\begin{figure}
	\includegraphics[width=\columnwidth]{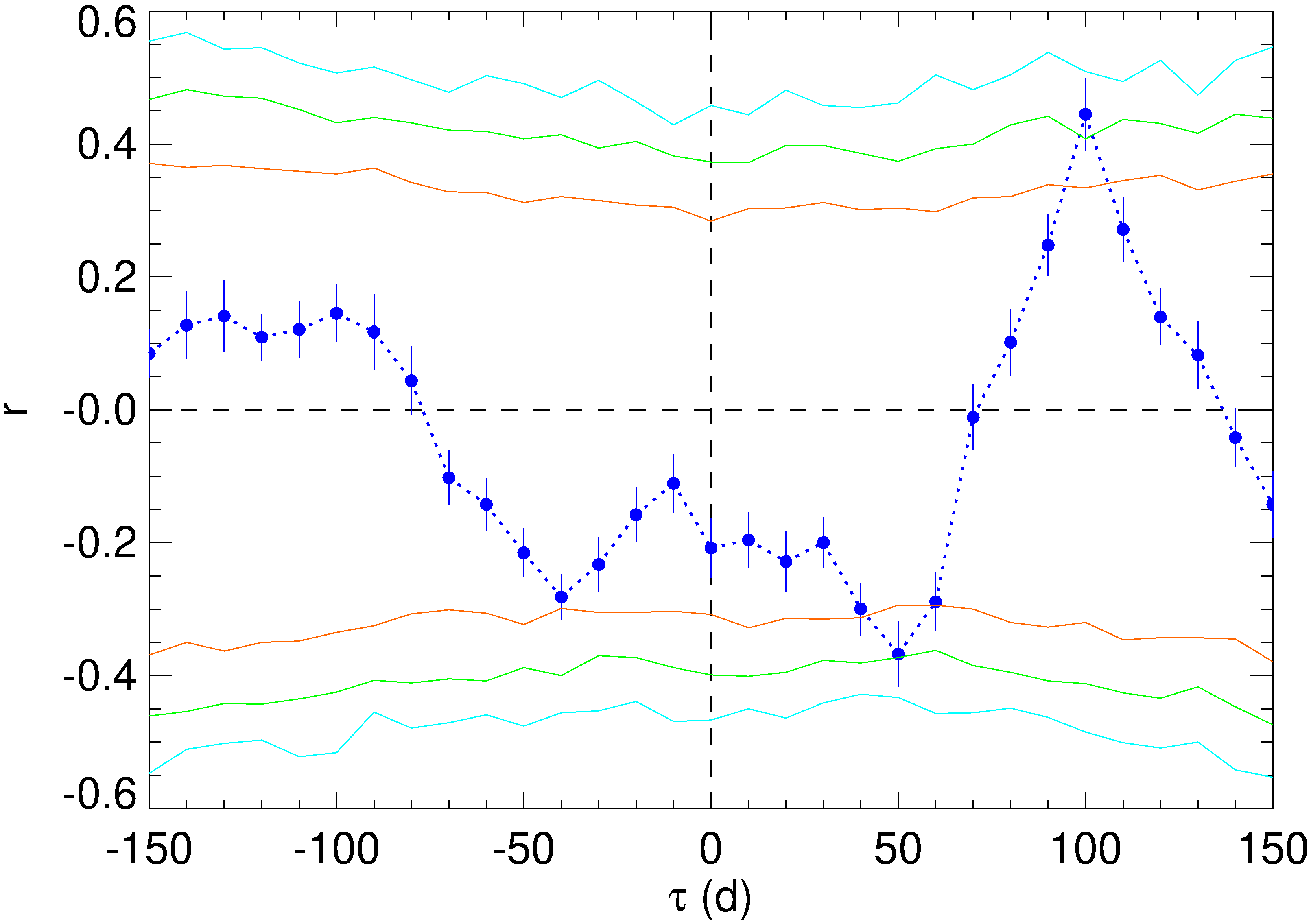}
    \caption{DCF between the $R$-band flux densities and the polarisation degree, with data binning of 2 days and DCF binnning of 10 days.
     The red, green, and cyan lines show the 80\%, 90\%, and 95\% confidence levels, respectively. These have been obtained through Monte Carlo simulations, as explained in the text.}
    \label{fig:dcf_lt}
\end{figure}

Fig.~\ref{fig:dcf_lt} shows the DCF between $F_R$ and $P$ for the whole period considered in this paper, adopting a data binning of 2 days and a DCF bin of 10 days. 
To estimate the significance of the DCF signals, we performed Monte Carlo simulations according to \citet{emma2013} and \citet{max2014}. We produced 1000 artificial time series with the same power spectral density (PSD) and probability density function (PDF) properties as the flux densities, and other 1000 for the polarisation degree, and then cross-correlated them.
The resulting 80\%, 90\%, and 95\% confidence levels are displayed in Fig.~\ref{fig:dcf_lt}.

The only noticeable peak, which, however, has a correlation coefficient $r$ of only $\sim 0.44$ and does not exceed the 95\% confidence level, suggests that $P$ follows $F_R$ with a delay of 100 days. This is the time interval between the flux major outburst phase around JD = 2459430 and the high level of $P$ reached about 100 days later and we do not attribute physical meaning to this signal. 
Instead, the anticorrelation at short time lags confirms what we have observed by visually inspecting Fig.~\ref{fig:pola}. As already mentioned there, this anticorrelation can be explained in terms of orientation changes of the optical-emitting jet region given a low value of the Lorentz factor, as detailed in \citet{raiteri2013}.

\begin{figure}
	% To include a figure from a file named example.*
	% Allowable file formats are eps or ps if compiling using latex
	% or pdf, png, jpg if compiling using pdflatex
	\includegraphics[width=\columnwidth]{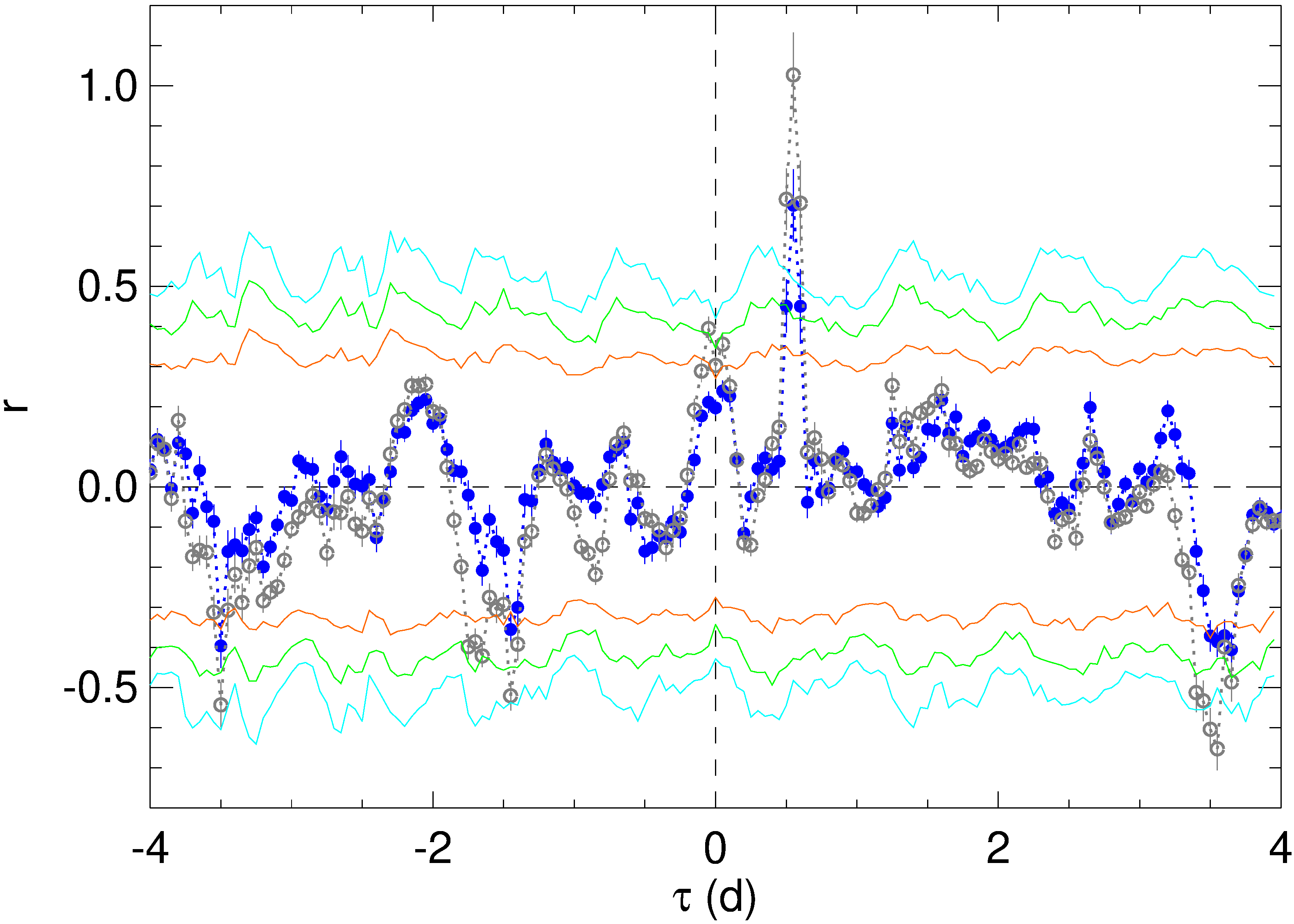}
    \caption{DCF between the deboosted $R$-band flux densities and the polarisation degree (blue dots). Data have been binned over 7.2 min intervals, and the DCF bin is 1.2 h. As in Fig.~\ref{fig:dcf_lt}, the red, green, and cyan lines show the 80\%, 90\%, and 95\% confidence levels, respectively. The grey empty circles represent the results of the DCF between the same deboosted $R$-band flux densities and the polarisation degree after removal of the long-term trend, showing how detrending makes signals become stronger.}
    \label{fig:dcf}
\end{figure}

We then analyse the correlation between the deboosted flux densities, i.e.\ what we consider the intrinsic jet flux variations, and $P$ with much finer binning. 
The result of the DCF between the deboosted flux densities and the polarization degree is shown in Fig.~\ref{fig:dcf}.
Data were preliminary binned over 7.2 minutes (0.005 days) and the DCF bin is 1.2 hours (0.05 days).
A noticeable peak whose value of the correlation coefficient is about 0.70 is found with a time delay of approximately 13 hours. This signal gets enhanced if we remove the long-term trend from the polarisation degree, dividing $P$ by the cubic spline interpolation shown in Fig.~\ref{fig:pola}. In this case the DCF peak has a value of 1.03. This means a strong correlation with a time delay of the polarisation degree variations with respect to the intrinsic flux changes. 
We notice that the same time lag of about 0.55 days of the $P$ variations after those of the flux density was also present in the past observing season according to the ZDCF analysis performed by \cite{jorstad2022}, although the correlation peak was lower. This indicates a persistent feature in the BL Lacertae behaviour.

To verify the robustness of the correlation signal, we estimated the significance of the DCF peak with the same method described above.
The corresponding confidence levels confirm the significance of the correlation with time lag of 0.55 days.

\begin{figure}
	% To include a figure from a file named example.*
	% Allowable file formats are eps or ps if compiling using latex
	% or pdf, png, jpg if compiling using pdflatex
	\includegraphics[width=\columnwidth]{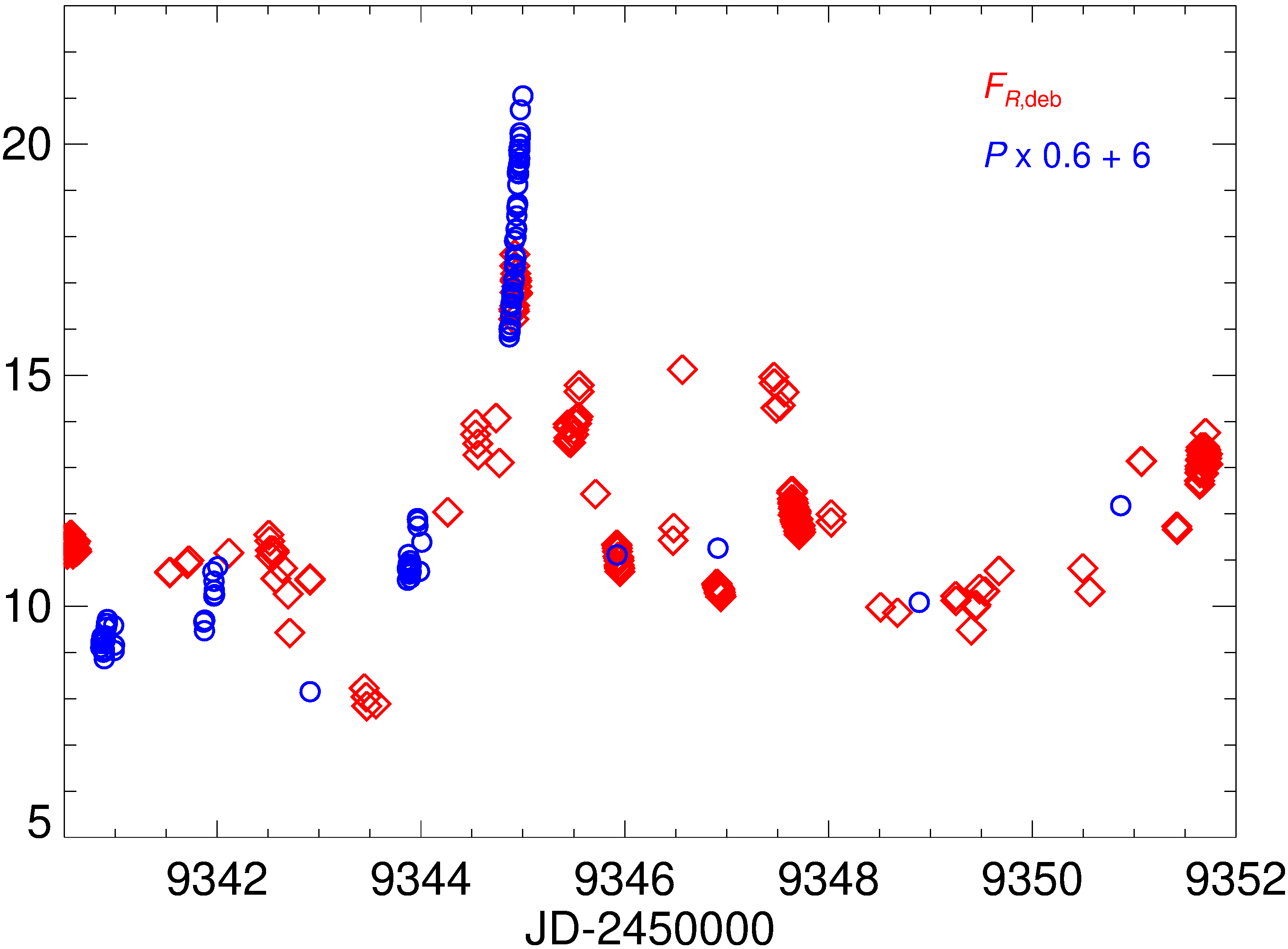}\\
        \vspace{0.1cm}\\
	\includegraphics[width=\columnwidth]{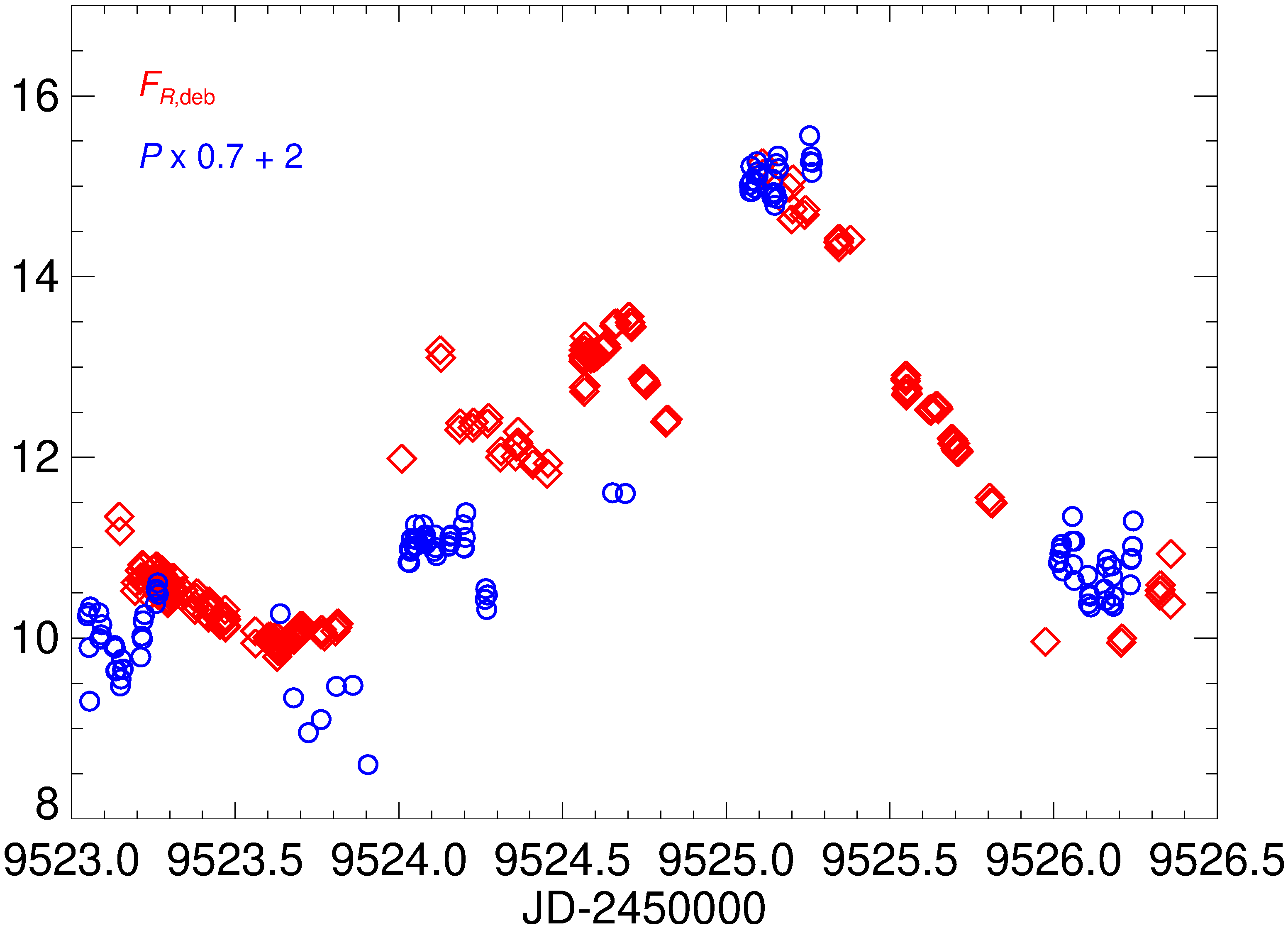}\\
    \caption{Deboosted $R$-band flux density (red diamonds) and polarisation degree (blue circles) in two time intervals with good polarisation sampling. $P$ is rescaled as indicated in the legenda and shifted in time by -0.55 days.}
    \label{fig:rp}
\end{figure}

This DCF signal actually comes from only two events, with $P$ peaking at $\rm JD \sim 2459345.6$ (2021 May 11) and $\rm JD \sim 2459525.8$ (2021 November 7), which have good sampling in both flux density and polarisation degree.
They are shown in Fig~\ref{fig:rp}, where $P$ is rescaled and shifted in time by -0.55 days in order to better match the deboosted flux densities. 
The reasonably close overlap of $F_R$ and $P$ further indicates that the corresponding flares have roughly the same duration in time.
%and strengthens the idea of a common nature.

During these events, the source flux density was approximately at the same low level, and the EVPA remained almost constant at about $101\degr$ and $19\degr$, respectively. This implies that the magnetic field in the first event was almost aligned with the direction of the jet as seen in the VLBA images at 43 GHz, while in the second event it was almost perpendicular \citep[see][]{jorstad2022}.
%$280\degr$ (equivalent to $100\degr$) and about $-160\degr$ (equivalent to $20\degr$), respectively.

The strong correlation between the jet intrinsic flux variations (deboosted flux densities) and the variations of the polarisation degree suggests that the same physical mechanism is responsible for both of them.
%The time delay of $P$ would most likely represent the time needed to reorder the magnetic field, after that some mechanism energized the particles and disrupted the magnetic field lines.

%We note that another, smaller peak is present in Fig.~\ref{fig:dcf} at zero time lag. 

%These results are compatible with the occurrence of QPOs with period of about 0.55 days in both flux and polarisation found by \citet{jorstad2022} in the 2020 observing season. A periodicity analysis with both wavelet and redfit methods in the time span covered by this paper did not lead to significant signals. However, we believe that the mechanism producing the 13 hours time scale unveiled in 2020 is still running and gives the DCF peaks at 0 and 13 hours.

%As for the other variations in deboosted flux and detrended polarisation degree, the polarimetric sampling is not dense enough to allow us to recognise possible further correlated events. In any case, it is generally accepted (ref) that there is an important component of the magnetic field affected by turbulence so that we must expect stochastic variations of flux and $P$. 

\section{Discussion and conclusions}
\label{sec:fine}
In this paper we have analysed the results of the optical photometric and polarimetric monitoring of BL Lacertae by the WEBT Collaboration in the 2021-2022 observing season.
A very dense sampling was achieved thanks to the common observing effort of many observers using 41 telescopes around the northern hemisphere.
In this period, the source showed intense variability, with continuous intraday flux variations, and reached its historical brightness maximum.
We have found evidence that the optical behaviour is well described by the twisting jet model, proposed in our earlier works to interpret the variability of BL Lacertae and other blazars.
According to our interpretation, the long-term variability is due to geometrical reasons, i.e., to the variation of the orientation of the optical-emitting jet region with respect to the line of sight. 

We have found several observational evidences supporting this view (see Sect.~\ref{sec:snake}), including enhanced amplitude variability during outbursts, and an almost achromatic long-term spectral trend.
In particular, by means of a wavelet analysis we have seen that significant variability short time scales appear only when the source is in outburst. The shortest time scales, from 2 days down to a few hours, are particularly evident in the brightest states. This is likely a consequence of the %time-scale reduction produced by the 
relativistic Doppler effect.
Indeed, these brightest states would be due to a high Doppler factor, which would also shorten the intrinsic time scales, leading to the appearance of the QPOs. 
%As a counter test, we can perform a wavelet analysis on the optical light curve corrected for the relativistic effects, where the flux densities are deboosted and the time intervals are deshortened using the variable Doppler factor estimated from the long-term trend. The results are shown in Fig.~\ref{fig:wavelet_deb2}, where the power appears much more homogeneous over the whole period, and the existence of the QPOs during outbursts is smoothed out. 
%We note that the QPOs detected in \citet{jorstad2022} occurred during a bright and constant state, which means a high and constant Doppler factor in our interpretation. The observed quasi-period of 0.55 days could be variable with brightness.
%\begin{figure*}
% \includegraphics[width=15cm]{plot_0p05_deb2_2021.png}
% \caption{As in Fig.~\ref{fig:wavelet}, but for the $R$-band light curve corrected for the relativistic effects on both flux and time as explained in the text.}
 %Results of the wavelet analysis. a) $R$-band flux densities of BL Lacertae in the 2021-2022 observing season. b) Wavelet power spectrum: the strength of the power is colour-coded according to the underlying palette; black contours define regions with confidence greater than 99\%; the grey grid represents the ``cone of influence" affected by edge effects. c) Global wavelet spectrum with 99\% confidence level marked by a dashed line. d) Time series of averaged periods between 0.1 and 2 days, with 99\% confidence level represented by the dashed line.}
% \label{fig:wavelet_deb2}
%\end{figure*}
Moreover, also the anti-correlation between the long-term variations in flux densities and degree of polarisation can be explained in the framework of our jet model, as discussed in \citet{raiteri2013}. 

Once the flux densities are corrected for the effects of the variable Doppler boosting, we are left with fast flux changes of roughly the same amplitude.
We recognize this short-term variability as the signature of energetic processes occurring inside the jet. 
In Sect.~\ref{sec:corre} we have found that variations in the deboosted flux densities are correlated with those in the degree of polarisation, which are following with a 0.55 day delay.
This correlation is based on a couple of well-sampled events, but was also present - though weaker - in the previous observing season \citep{jorstad2022}, suggesting a common feature in the source behaviour.
Interestingly, 0.55 day is also the period of the QPOs that \citet{jorstad2022} detected in the optical flux and polarisation, and in the $\gamma$-ray flux of BL Lacertae during the first phase of the 2020 outburst. In that case, the QPOs were explained as the result of a current-driven kink instability that was triggered by the passage of an off-axis perturbation through a stationary recollimation shock.

Magnetohydrodynamic simulations of relativistic plasma jets (RMHD) show that jets are subject to kink instabilities that twist the magnetic field lines and can disrupt the magnetic field structure. Magnetic reconnection then reorganizes the field, dissipating magnetic energy into particle energy \citep{begelman1998,sironi2015,zhang2018,dong2020,zhang2020,acharya2021,bodo2021,zhang2022}. The emission would come from plasmoids, dynamic plasma blobs/magnetic structures that appear in the reconnection region and that can have up to 10\% of its size \citep[e.g][]{petropoulou2018,zhang2022}.
In simulations including polarisation, flux peaks correspond to low $P$ states, while $P$ increases when the flux decreases, and the EVPA can either undergo large rotations or remain roughly constant. 
According to RMHD simulations by \citet{dong2020}, kink instabilities lead to a quasi-periodic release of energy.
%, with period determined by the kink growing time. 
QPOs result also from the RMHD simulations by \citet{acharya2021}, who confirmed the validity of the twisting jet model.
These kink-driven QPOs could explain the short time scales appearing in the optical light curve of BL Lacertae during the brightest states (see Sect.~\ref{sec:wavelet}).
In this view, the QPOs time scales could be linked to the dimension of the plasmoids, and the delay between the flux and $P$ variations could represent the time needed to recover an ordered magnetic field, after its disruption by turbulence driven by the dissipation episode that energised the particles.

%The effects of magnetic reconnection triggered by current-driven kink instabilities were studied by \citet{bodo2021} in the framework of RMHD simulations. These show that during an emission flare $P$ remains low, while in the post-flare phase, $P$ can undergo noticeable fluctuations. This is likely because dissipation due to magnetic reconnection energizes particles giving rise to the flux increase, and triggers turbulence, which destroys the magnetic field and makes $P$ decrease. Then the flux fades and the magnetic field comes back to a more ordered situation, so that $P$ increases. 

An alternative explanation for the correlation between $F$ and $P$ that we observed involves shock waves \citep[e.g.][]{marscher1985,spada2001,sironi2015,boettcher2019}. Shocks could propagate through the optical emitting region, promptly accelerating particles and gradually ordering the magnetic field. 
One caveat comes from the fact that during the first of the two events 
producing the correlation,
the EVPA was about 100\degr, which means that the magnetic field was almost aligned with the radio jet. In contrast, during the second event the EVPA was about 20\degr, i.e.\ the magnetic field was approximately perpendicularly to the radio jet. 
If we assume a straight jet, it is difficult to imagine that these different situations can lead to the same lag between the variations in $F$ and $P$. 
If we instead assume an inhomogeneous twisting jet, the jet direction in the optical emitting region varies continuously as the jet twists, and it is likely different from the apparent orientation of the observed radio jet.
The two events occurred in similar low brightness states, 
%F~45 mJy nel primo evento, F~50 nel secondo
which in our geometrical interpretation implies that the Doppler beaming was similar and low, i.e., in both cases the optical emitting region had a similar viewing angle, but not necessarily the same orientation in space. Indeed, between the two events there is a complete rotation of the EVPA of more than 360\degr, which strengthens the idea of a twisting jet.
In our view it is the jet that changes orientation, and we can have two different observed values of the EVPA in the optical emitting region while the orientation of the magnetic field with respect to the jet direction remains constant. This could explain why shocks can produce similar events with different observed EVPAs.

%As a final remark, we recall that any time scale in fast variations or QPO is affected by the shortening caused by the relativistic Doppler effect. Therefore, any search for QPOs must take into account the variable brightness level due to Doppler boosting. If a typical time scale exists in the jet rest frame, it will appear shorter in brighter states than in fainter ones. One possibility is to inspect only fairly constant brightness levels, as e.g. done in Jorstad et al. The other is to deal with a light curve with fluxes corrected for the Doppler boosting and time intervals corrected for the same Doppler changes. The same argument can be applied to the time delay of P versus F, which actually we found to be 0.55 days just in similar brightness levels.

A final remark is due regarding the consequences of the relativistic effects on the search for QPOs, since changes in the Doppler factor affect not only the source brightness, but also the time scales of the intrinsic, energetic variations. If a typical time scale exists in the jet rest frame, it will appear shorter in brighter states than in fainter ones. Therefore, the search of QPOs is meaningful only when performed during fairly constant brightness levels, as e.g. done in \citet{jorstad2022}. Otherwise, one must process the light curves in order to correct both the fluxes and the time intervals for the same Doppler factor changes. This argument can also be applied to the time delay of $P$ versus $F$, and indeed we found the same delay of 0.55 days in similar brightness levels.

%\citet{hazama2022} analysed the BL Lacertae optical behaviour in the first part of the period covered by this paper, although with much less sampling. They suggested that the high state occurring at $\rm JD \sim 2459248$ could be caused by a shock produced by the collision of plasma blobs, since the magnetic field was perpendicular to the jet direction. In contrast, the outburst at $\rm JD \sim 2459433$, characterised by $P$ decreasing with increasing flux and a rapidly variable EVPA, is possibly due to variation of the jet viewing angle and magnetic reconnection,

%The low $P$ phases can be explained through turbulence, which destroys the average magnetic field and produces random flux and EVPA variations.

%A time scale of 13 hours implies a dimension of the jet region where the phenomenon occurs of the order of $r = t \, c \, \delta / (1+z) \sim 1.3 \, 10^{16} \rm \, cm \sim 0.004 \, pc$, for a Doppler factor $\delta=10$.

%La dissipazione energizza le particelle, crea turbolenza  e distrugge quindi il campo magnetico, per cui aumenta il flusso e scende il grado di polarizzazione. Quando il fenomeno si spegne, il flusso cala e il campo magnetico ritorna ordinato.
%La kink instability crea un getto elicoidale. Se ci sono tante spire che emettono o tante zone che si accendono, mediamente l'effetto si smorza; se pero` una predomina, si puo` vedere un effetto di quel tipo. Vedere Bodo, Tavecchio e Sironi.

\section*{Acknowledgements}
We acknowledge useful discussions with Haocheng Zhang, Gianluigi Bodo and Paola Rossi.
We are grateful to Massimo Conti, Paolo Rosi e Luz Marina Tinjaca Ramirez for help with the observations at the Montarrenti Observatory.
Wavelet software was provided by C.~Torrence and G.~Compo, and is available at URL: http://paos.colorado.edu/research/wavelets/.
Light curve simulations were based on the S.~D.~Connolly python version of the \citet{emma2013} algorithm, which is available at https://github.com/samconnolly/DELightcurveSimulation \citep{connolly2015}.
This research has made use of NASA’s Astrophysics Data System Bibliographic Services.
The Torino group acknowledges contribution from the grant INAF Main Stream project ``High-energy extragalactic astrophysics: toward the Cherenkov Telescope Array".
The BU group was supported in part by U.S. National Science Foundation grant AST-2108622 and NASA Fermi GI grant 80NSSC22K1571. This study was based (in part) on observations conducted using the 1.8 m Perkins Telescope Observatory (PTO) in Arizona (USA), which is owned and operated by Boston University.
The Connecticut College group thanks the College's Research Matters fund for partially supporting the acquisition of our observations. 
Based on observations made with the Nordic Optical Telescope, owned in collaboration by the University of Turku and Aarhus University, and operated jointly by Aarhus University, the University of Turku and the University of Oslo, representing Denmark, Finland and Norway, the University of Iceland and Stockholm University at the Observatorio del Roque de los Muchachos, La Palma, Spain, of the Instituto de Astrofisica de Canarias.
This paper is partly based on observations made with the IAC-80 telescope operated on the island of Tenerife by the Instituto de Astrofísica de Canarias in the Spanish Observatorio del Teide and on observations made with the LCOGT 0.4 m telescope network, one of whose nodes is located in the Spanish Observatorio del Teide. 
This research was partially supported by the Bulgarian National Science Fund of the Ministry of Education and Science under grants KP-06-H28/3 (2018), KP-06-H38/4 (2019), KP-06-KITAJ/2 (2020) and KP-06-PN68/1(2022). The Skinakas Observatory is a collaborative project of the University of Crete, the Foundation for Research and Technology -- Hellas, and the Max-Planck-Institut f\"ur Extraterrestrische Physik.
We acknowledge support by Bulgarian National Science Fund under grant DN18-10/2017 and National RI Roadmap Project D01-176/29.07.2022 of the Ministry of Education and Science of the Republic of Bulgaria.
K.M. acknowledges support for the Osaka observations by the JSPS KAKENHI grant No. 19K03930.
GD, OV, MS, and MDJ acknowledge the observing grant support from the Institute of Astronomy and Rozhen NAO BAS through
the bilateral joint research project ``Gaia Celestial Reference Frame (CRF) and the fast variable astronomical objects" (2020-2022, leader is G.Damljanovic), and support by the Ministry of Education, Science and Technological Development of the Republic of Serbia (contract No 451-03-68/2022-14/200002).
JOS thanks the support from grant FPI-SO from the Spanish Ministry of Economy and Competitiveness (MINECO) (research project SEV-2015-0548-17-3 and predoctoral contract BES-2017-082171). JOS and JAP acknowledge financial support from the Spanish Ministry of Science and Innovation (MICINN) through the Spanish State Research Agency, under Severo Ochoa Program 2020- 2023 (CEX2019-000920-S). 
The R-band photometric data from the University of Athens Observatory (UOAO) were obtained after utilizing the robotic and remotely controlled instruments at the facilities \citep{gazeas2016}.
ACG is partially supported by Chinese Academy of Sciences (CAS) President’s International Fellowship Initiative (PIFI) (grant no. 2016VMB073), Ministry of Science and Technology of China (grant No. 2018YFA0404601), and the National Science Foundation of China (grants No. 11621303, 11835009, and 11973033).  HG acknowledges financial support from the Department of Science and Technology (DST), Government of India, through INSPIRE faculty award IFA17-PH197 at ARIES, Nainital, India.  PK acknowledges support from ARIES A-PDF grant (AO/APDF/770), and support from the Department of Science and Technology (DST), Government of India, through the DST-INSPIRE faculty grant (DST/INSPIRE/04/2020/002586).
The Abastumani team acknowledges financial support by the Shota Rustaveli NSF of Georgia under contract FR-19-6174.
%This work is based upon observations carried out at the Observatorio Astronómico Nacional on the Sierra San Pedro Martir (OAN-SPM), Baja California, Mexico.  

%This research has made use of data from the OVRO 40-m monitoring program \citep{richards2011}, supported by private funding from the California Institute of Technology and the Max Planck Institute for Radio Astronomy, and by NASA grants NNX08AW31G, NNX11A043G, and NNX14AQ89G and NSF grants AST-0808050 and AST- 1109911.

%%%%%%%%%%%%%%%%%%%%%%%%%%%%%%%%%%%%%%%%%%%%%%%%%%
\section*{Data Availability}
    Data acquired by the WEBT collaboration are stored in the WEBT archive and are available upon request to the WEBT President Massimo Villata (\href{mailto:massimo.villata@inaf.it}{massimo.villata@inaf.it}).
%    The $\gamma$-ray light curves used for this paper have been downloaded from the {\it Fermi} LAT Light Curve Repository (\url{https://fermi.gsfc.nasa.gov/ssc/data/access/lat/LightCurveRepository/}).

%%%%%%%%%%%%%%%%%%%% REFERENCES %%%%%%%%%%%%%%%%%%

% The best way to enter references is to use BibTeX:

\bibliographystyle{mnras}
\bibliography{bllac_2021} % if your bibtex file is called example.bib

% Alternatively you could enter them by hand, like this:
% This method is tedious and prone to error if you have lots of references
%\begin{thebibliography}{99}
%\bibitem[\protect\citeauthoryear{Author}{2012}]{Author2012}
%Author A.~N., 2013, Journal of Improbable Astronomy, 1, 1
%\bibitem[\protect\citeauthoryear{Others}{2013}]{Others2013}
%Others S., 2012, Journal of Interesting Stuff, 17, 198
%\end{thebibliography}

%%%%%%%%%%%%%%%%%%%%%%%%%%%%%%%%%%%%%%%%%%%%%%%%%%

 \section*{Affiliations}
{\it
$^{ 1}$INAF, Osservatorio Astrofisico di Torino, via Osservatorio 20, I-10025 Pino Torinese, Italy                                                                                                             \\
$^{ 2}$Institute for Astrophysical Research, Boston University, 725 Commonwealth Avenue, Boston, MA 02215, USA                                                                                                 \\
$^{ 3}$Astronomical Institute, St. Petersburg State University, 28 Universitetskiy prospekt, Peterhof, St. Petersburg, 198504, Russia                                                                          \\
$^{ 4}$Instituto de Astrofísica de Canarias (IAC), E-38200 La Laguna, Tenerife, Spain                                                                                                                         \\
$^{ 5}$Universidad de La Laguna (ULL), Departamento de Astrofisica, E-38206, Tenerife, Spain                                                                                                                   \\
$^{ 6}$EPT Observatories, Tijarafe, La Palma, Spain                                                                                                                                                            \\
$^{ 7}$INAF, TNG Fundaci\'on Galileo Galilei, La Palma, Spain                                                                                                                                                  \\
$^{ 8}$Institute of Astronomy, National Central University, Taoyuan 32001, Taiwan                                                                                                                              \\
$^{ 9}$Department of Physics and Astronomy, N283 ESC, Brigham Young University, Provo, UT 84602, USA                                                                                                           \\
$^{10}$Abastumani Observatory, Mt. Kanobili, 0301 Abastumani, Georgia                                                                                                                                          \\
$^{11}$Hans-Haffner-Sternwarte, Naturwissenschaftliches Labor f\"ur Sch\"uler am FKG; Friedrich-Koenig-Gymnasium, 97082 W\"urzburg, Germany                                                                    \\
$^{12}$Astronomical Observatory, Department of Physical Sciences, Earth and Environment, University of Siena, Siena, Italy                                                                                     \\
$^{13}$Astronomical Institute, Osaka Kyoiku University, Kashiwara, Japan                                                                                                                                       \\
$^{14}$Ulugh Beg Astronomical Institute, Astronomy Street 33, Tashkent 100052, Uzbekistan                                                                                                                      \\
$^{15}$Special Astrophysical Observatory, Russian Academy of Sciences, 369167, Nizhnii Arkhyz, Russia                                                                                                          \\
$^{16}$Pulkovo Observatory, St. Petersburg, 196140, Russia                                                                                                                                                     \\
$^{17}$Institute of Astronomy and National Astronomical Observatory, Bulgarian Academy of Sciences, 72 Tsarigradsko shosse Blvd., 1784 Sofia, Bulgaria                                                         \\
$^{18}$Astronomical Observatory, Volgina 7, 11060 Belgrade, Serbia                                                                                                                                             \\
$^{19}$Osservatorio Astronomico Citt\`a di Seveso, Seveso, Italy                                                                                                                                               \\
$^{20}$Department of Aerospace Science and Technology, Politecnico di Milano, Milano, Italy                                                                                                                    \\
$^{21}$Gruppo Astrofili Catanesi (GAC), Catania, Italy                                                                                                                                                         \\
$^{22}$INAF, Osservatorio Astronomico di Brera, via E. Bianchi 46, I-23807 Merate, Italy                                                                                                                      \\
$^{23}$Crimean Astrophysical Observatory RAS, P/O Nauchny, 298409, Russia                                                                                                                                      \\
$^{24}$Department of Astronomy, Faculty of Physics, Sofia University ``St. Kliment Ohridski", 5 James Bourchier Blvd., BG-1164, Sofia, Bulgaria                                                                 \\
$^{25}$Department of Physics, Astronomy and Geophysics, Connecticut College, New London, CT 06320, USA                                                                                                         \\
$^{26}$Aryabhatta Research Institute of Observational Sciences (ARIES), Manora Peak, Nainital 263001, India                                                                                                    \\
$^{27}$School of Studies in Physics \& Astrophysics, Pt. Ravishankar Shukla University, Amanaka G.E. Road, Raipur 492010, India                                                                                \\
$^{28}$National University of Uzbekistan, Tashkent 100174, Uzbekistan                                                                                                                                          \\
$^{29}$Fakult\"at Physik, TU Dortmund, 44227 Dortmund, Germany                                                                                                                                                 \\
$^{30}$Hypatia Observatory, Rimini, Italy                                                                                                                                                                      \\
$^{31}$GiaGa Observatory, Pogliano Milanese, Italy                                                                                                                                                             \\
$^{32}$Section of Astrophysics, Astronomy and Mechanics, Department of Physics, National and Kapodistrian University of Athens, GR-15784 Zografos, Athens, Greece                                              \\
$^{33}$Key Laboratory for Research in Galaxies and Cosmology, Shanghai Astronomical Observatory, Chinese Academy of Sciences, Shanghai 200030, China                                                           \\
$^{34}$Shanghai Frontiers Science Center of Gravitational Wave Detection, 800 Dongchuan Road, Minhang, Shanghai 200240, China                                                                                  \\
$^{35}$Department of Physical Science, Aoyama Gakuin University, 5-10-1 Fuchinobe, Chuo-ku, Sagamihara-shi, Kanagawa 252-5258, Japan                                                                           \\
$^{36}$Department of Physics and Astronomy, Faculty of Natural Sciences, University of Shumen, 115, Universitetska Str., 9712 Shumen, Bulgaria                                                                 \\
$^{37}$Department of Physics, DDU Gorakhpur University, Gorakhpur 273009, India                                                                                                                                \\
$^{38}$Zentrum f\"ur Astronomie der Universit\"at Heidelberg, Landessternwarte, K\"onigstuhl 12, 69117 Heidelberg, Germany                                                                                     \\
$^{39}$Engelhardt Astronomical Observatory, Kazan Federal University, Tatarstan, Russia                                                                                                                        \\
$^{40}$Department of Physical Sciences, Indian Institute of Science Education and Research (IISER) Mohali, Knowledge City, Sector 81, SAS Nagar, Punjab 140306, India                                          \\
$^{41}$Abbey Ridge Observatory, Canada                                                                                                                                                                         \\
$^{42}$Montarrenti Observatory, Siena, Italy                                                                                                                                                                   \\
$^{43}$Lehrstuhl f\"ur Astronomie, Universit\"at W\"urzburg, 97074 W\"urzburg, Germany                                                                                                                         \\
$^{44}$Wild Boar Remote Observatory, Florence, Italy                                                                                                                                                           \\
$^{45}$Crimean Astrophysical Observatory of the Russian Academy of Sciences, P/O Nauchny, 298409, Russia                                                                                                       \\
$^{46}$Nordic Optical Telescope Apartado 474, E-38700 Santa Cruz de La Palma, Santa Cruz de Tenerife, Spain                                                                                                    \\
$^{47}$Department of Physics and Astronomy, Aarhus University, Munkegade 120, DK-8000 Aarhus C, Denmark                                                                                                        \\
$^{48}$University of Craiova, Alexandru Ioan Cuza 13, 200585 Craiova, Romania                                                                                                                                  \\
$^{49}$AAVSO observer, Russia                                                                                                                                                                                  \\
$^{50}$Remote observer of Burke-Gaffney Observatory and Abbey Ridge Observatory, Canada                                                                                                                        \\
 }
 %%%%%%%%%%%%%%%%% APPENDICES %%%%%%%%%%%%%%%%%%%%%

%\appendix

%\section{Some extra material}

%If you want to present additional material which would interrupt the flow of the main paper,
%it can be placed in an Appendix which appears after the list of references.

%%%%%%%%%%%%%%%%%%%%%%%%%%%%%%%%%%%%%%%%%%%%%%%%%%

% Don't change these lines
\bsp	% typesetting comment
\label{lastpage}
\end{document}